\documentclass[12pt,a4paper]{article}
\pdfoutput=1
\usepackage{epsfig,amssymb,amsmath,graphicx,caption,subcaption,verbatim,hyperref,xcolor,ulem,epstopdf,psfrag,pstool,braket,array,enumerate,wrapfig,tabularx,wrapfig}
\usepackage{graphics,color, ltablex,booktabs}
\usepackage[font={small}]{caption}
\numberwithin{equation}{section}

\begin{document} 

\newcommand{\be}{\begin{equation}}
\newcommand{\ee}{\end{equation}}
\newcommand{\bea}{\begin{eqnarray}}
\newcommand{\eea}{\end{eqnarray}}
\newcommand{\bean}{\begin{eqnarray*}}
\newcommand{\eean}{\end{eqnarray*}}
\font\upright=cmu10 scaled\magstep1
\font\sans=cmss12
\newcommand{\ssf}{\sans}
\newcommand{\stroke}{\vrule height8pt width0.4pt depth-0.1pt}
\newcommand{\Z}{\hbox{\upright\rlap{\ssf Z}\kern 2.7pt {\ssf Z}}}
\newcommand{\ZZ}{\Z\hskip -10pt \Z_2}
\newcommand{\C}{{\rlap{\upright\rlap{C}\kern 3.8pt\stroke}\phantom{C}}}
\newcommand{\R}{\hbox{\upright\rlap{I}\kern 1.7pt R}}
\newcommand{\HH}{\hbox{\upright\rlap{I}\kern 1.7pt H}}
\newcommand{\CP}{\hbox{\C{\upright\rlap{I}\kern 1.5pt P}}}
\newcommand{\identity}{{\upright\rlap{1}\kern 2.0pt 1}}
\newcommand{\half}{\frac{1}{2}}
\newcommand{\quart}{\frac{1}{4}} 
\newcommand{\pr}{\partial}
\newcommand{\bm}{\boldmath}
\newcommand{\I}{{\cal I}} 
\newcommand{\M}{{\cal M}}
\newcommand{\N}{{\cal N}}
\newcommand{\e}{\varepsilon}
\newcommand{\ep}{\epsilon}
\newcommand{\balpha}{\mbox{\boldmath $\alpha$}}
\newcommand{\bgamma}{\mbox{\boldmath $\gamma$}}
\newcommand{\blambda}{\mbox{\boldmath $\lambda$}}
\newcommand{\bep}{\mbox{\boldmath $\varepsilon$}}
\newcommand{\Oh}{{\rm O}}
\newcommand{\x}{{\bf x}}
\newcommand{\y}{{\bf y}}
\newcommand{\bR}{{\bf R}}
\newcommand{\bl}{{\bf l}}
\newcommand{\bJ}{{\bf J}}
\newcommand{\X}{{\bf X}}
\newcommand{\Y}{{\bf Y}}
\newcommand{\z}{{\bar z}}
\newcommand{\w}{{\bar w}}
\newcommand{\tT}{{\tilde T}}
\newcommand{\tX}{{\tilde\X}}
\def\ir3{\int_{\mathbb{R}^{3}}}

\thispagestyle{empty}
\rightline{DAMTP-2019-6}
\begin{center}
{{\bf \Large Oxygen-16 Spectrum from Tetrahedral Vibrations
and their Rotational Excitations}} 
\\[8mm]

{\bf \large C.~J. Halcrow\footnote{email: C.J.Halcrow@leeds.ac.uk}} \\[1pt]
{\it 
School of Mathematics, University of Leeds,\\
Leeds LS2 9JT, U.K.} \\[20pt]

{\bf \large C. King\footnote{email: ck402@cam.ac.uk} 
and N.~S. Manton\footnote{email: N.S.Manton@damtp.cam.ac.uk}} \\[1pt]
{\it 
Department of Applied Mathematics and Theoretical Physics,\\
University of Cambridge,
Wilberforce Road, Cambridge CB3 0WA, U.K.}
\vspace{1mm}

\end{center}

\abstract{}
    
A reinterpretation of the complete energy spectrum of the Oxygen-16 nucleus
up to 20 MeV, and partly beyond, is proposed. The underlying intrinsic 
shape of the nucleus is tetrahedral, as in the na\"ive alpha-particle 
model and other cluster models, and A, E and F vibrational phonons 
are included. The A- and F-phonons are treated in the harmonic
approximation, but the E-vibrations are extended into a
two-dimensional E-manifold of $D_2$-symmetric, four-alpha-particle 
configurations, following earlier work. This allows for the underlying 
tetrahedral configuration to tunnel through a square configuration 
into the dual tetrahedron, with the associated breaking of parity 
doubling. The frequency of an E-phonon is lower than in
other models, and the first-excited $0^+$ state at 6.05 MeV is
modelled as a state with two E-phonons; this allows a good fit of the 
lowest $2^+$ and $2^-$ states as excitations with one E-phonon. 
Rotational excitations of the vibrational states are analysed as in 
the classic work of Dennison, Robson and others, with centrifugal 
corrections to the rotational energy included. States with F-phonons require 
Coriolis corrections, and the Coriolis parameter $\zeta$ is chosen 
positive to ensure the right splitting of the $3^+$ and $3^-$ states near 
11 MeV. Altogether, about 80 states with isospin zero are predicted below 
20 MeV, and these match rather well the more than 60 experimentally 
tabulated states. Several high-spin states are predicted, up to spin 
9 and energy 30 MeV, and these match some of the observed high-spin, 
natural parity states in this energy range. The model proposed here is 
mainly phenomenological but it receives some input from analysis of 
Skyrmions with baryon number 16.
  
\vskip 5pt

\vfill
\newpage
\setcounter{page}{1}
\renewcommand{\thefootnote}{\arabic{footnote}}


\section{Introduction} 

For many decades there have been competing views of the intrinsic
structure of the Oxygen-16 nucleus. In the na\"ive alpha-particle model,
the nucleus is a cluster structure of four alpha particles at the 
vertices of a regular tetrahedron \cite{Wheel}. The binding energy of
Oxygen-16 and of several other small nuclei that contain a whole
number of alpha particles can be interpreted in terms of the number of 
short bonds between them \cite{Wef,HT}; for Oxygen-16 there are six 
bonds. In the shell model, on the other hand, there is an 
underlying spherically-symmetric potential, and the eight protons and 
eight neutrons fill all the available, lowest-lying 1s- and 1p-states,
making the nucleus as a whole spherical, and magic. However, it is 
well known that the shell model picture is not completely 
incompatible with alpha-particle clustering in the Oxygen-16 ground 
state, and some extent of alpha clustering is confirmed in many 
experiments \cite{SkPerr,VT}. For a review of alpha-particle 
clustering in light nuclei, see \cite{Fuj}, and for recent discussions 
of the cluster structure in Oxygen-16, see \cite{K-E,FHK-E,FK-E}. 
Some insight is afforded by {\it ab initio} calculations involving 16 
nucleons with realistic 2-body and 3-body forces \cite{abin}. 
Tetrahedral and square clusters of alpha particles appear to be 
favoured. 

Of course, as the ground state of Oxygen-16 has $J^P=0^+$, the mean
particle density in this state is spherical in any model. However, 
a conceptual difference arises for the low-lying $3^-$ state at 6.13
MeV, which is known through its E3 decay strength to be a highly collective 
excitation. If the ground-state intrinsic shape is spherical, then
this state is a vibrational excitation, perhaps with a tetrahedral 
character to account for the spin and parity. If the intrinsic shape 
is tetrahedral then this state is simply a rotational excitation. 

The real challenge is to understand not just the ground state and
a few low-lying excitations, but the entire known spectrum of 
excited states of Oxygen-16. A significant number of states can be
described in terms of particle-hole excitations within the shell
model. This was systematized by Brink and Nash \cite{BrinkShell}; see also 
D. J. Millener's theoretical foreword in \cite{O16Exp}. The single particle-hole states all have negative parity and describe a portion of the low-lying spectrum. To accurately model the known experimental energies, one must use techniques going beyond the independent particle version of the shell model, like the Tamm--Dacoff approach and the RPA \cite{Suh}. 
Even-parity states, including the $0^+$ state at 6.05 MeV are
less easily described in the shell model. Some of these states 
can be modelled as correlated 4-particle, 4-hole 
states \cite{4h4p,review}, which can 
be interpreted as states with a mobile alpha particle that 
changes the underlying spherical shape. However, a systematic
study of all excited states within the shell model would require a 
prohibitively large number of higher-energy 1-particle states to be 
activated, beyond the sd-shell. This leads us back to cluster models 
and collective excitations.

Our proposal, then, builds on the long history of modelling the excitation
spectrum of Oxygen-16 in terms of vibrational excitations of a
tetrahedral intrinsic structure. The many studies of possible
tetrahedral structures in larger nuclei provide further 
stimulus \cite{TYM,DGSM,TSD,Man,Heu}. Each vibrational state has an
associated rotational band of rovibrational states, where the allowed 
spin/parities are controlled by the representation theory of the 
tetrahedral group. The earliest work on this, following Wheeler \cite{Wheel},
appears to be that of Dennison \cite{Denn}, who applied to the
Oxygen-16 nucleus many insights gained from
studying the spectra of tetrahedral molecules like methane 
(${\rm CH}_4$). In particular, it was known that a tetrahedral structure of
four alpha particles has three vibrational frequencies, associated
with irreducible representations (irreps) $A$, $E$ and $F$ of the 
tetrahedral group\footnote{We shall be
more careful later to distinguish the $A_1$, $A_2$, $E$, $F_1$ and
$F_2$ irreps of $T_d$.}. These irreps have,
respectively, degeneracies 1, 2 and 3. Also known was that in the
rotational excitations of a state with one vibrational F-phonon, it is
important to take account of the (quantum) Coriolis effect.
Dennison's work was followed up by Kameny \cite{Kam} and then by
Robson \cite{Rob}. By Robson's time, around 1980, the experimental
states were much better known, so parameters like the vibrational 
frequencies and moments of inertia were better understood. Robson's fit of 
the spectrum led to a prediction of a $3^-$ state at 9.85 MeV that had not
been seen. An experiment shortly afterwards at Florida State University
confirmed that there was no such state \cite{Fra}, which was a great
disappointment, and struck a blow for this approach. However, the blow
is not fatal, as we will show. The rovibrational model was revived by 
Bauhoff, Schultheis and Schultheis \cite{Bau}, who found rotational 
bands for a number of different alpha-particle configurations,
including tetrahedral, kite and chain clusters. These clusters are 
local minima of an energy function constructed from nucleon-nucleon 
forces. More recently, Bijker and Iachello \cite{TetVib} explored the
consequences of novel, larger symmetry
algebras that could predict the relative frequencies of the A, E and F
vibrational modes. In particular, they considered the possibility that these
frequencies are all equal at about 6 MeV, and they calculated the
rotational excitations, successfully fitting about 10 known states. As
in the earlier work, they drew attention to Coriolis contributions to
the energy. However, there are clearly limitations to the success 
of their fit. An alternative rovibrational model is needed.

\begin{figure}[h!]
	\begin{center}
		\includegraphics[width=0.6\textwidth]{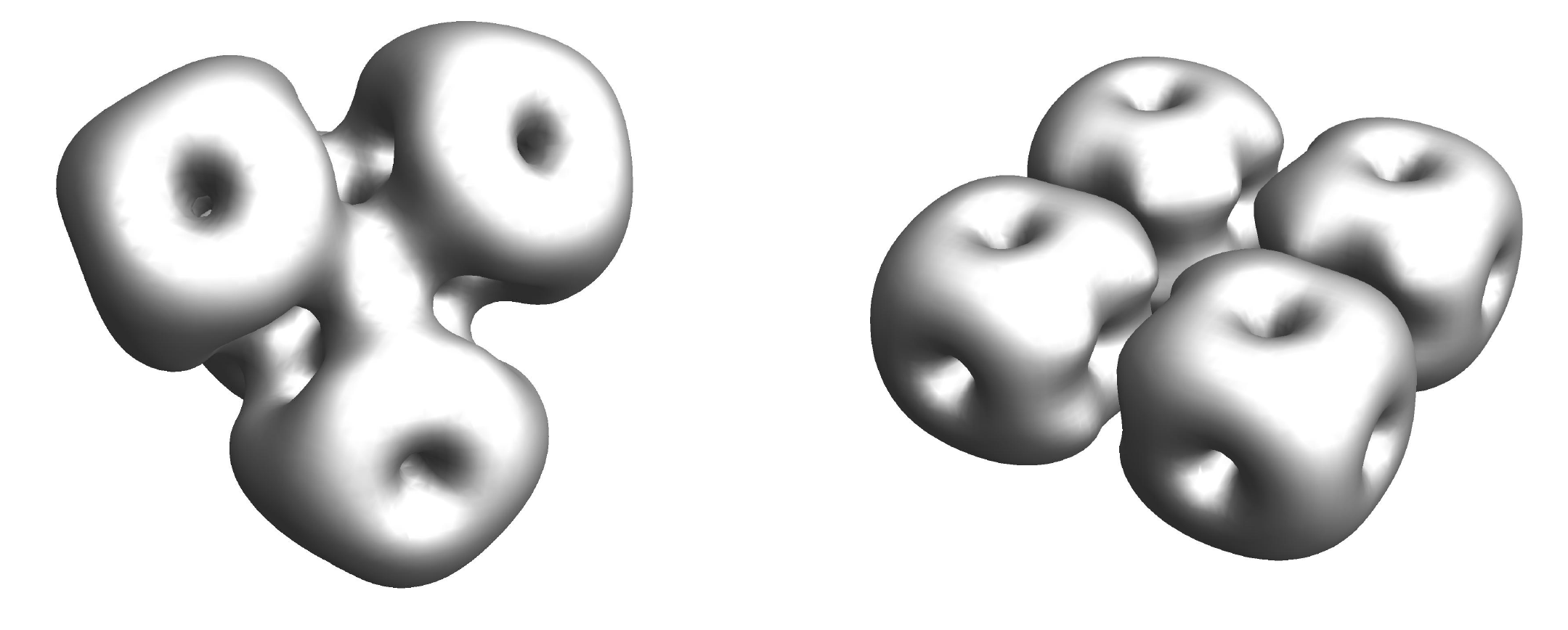} 
	\end{center}
	\caption{ Tetrahedral and square Skyrmion configurations with
baryon number 16. A contour of constant energy density of the Skyrme 
field is shown. The constituent cubes can be interpreted as alpha
particles. } \label{Skyrmions}
\end{figure}

We previously considered the Oxygen-16 spectrum in \cite{HKM}. Our work
developed from studies of Skyrmions, the solitons in an effective
field theory of pions, whose topological charge is identified with
baryon number (equivalently, atomic mass number) \cite{Sky}. In 
the Skyrme model a solution with baryon number 16, which is the basis 
for modelling Oxygen-16, has tetrahedral symmetry \cite{BMS,LM2}. We
will not review Skyrmions in any detail here, as our rovibrational model for 
Oxygen-16 hardly depends on any variant of Skyrme's field theory. The main
insight we have gained from Skyrmions is that in addition to the
solution with baryon number 16 having tetrahedral symmetry, there is 
another solution of comparable energy with square symmetry, see 
Fig. \ref{Skyrmions} \cite{BMS}. The first solution mimics 
four alpha particles arranged as a tetrahedron, and the second, four 
alpha particles arranged as a square; the alpha particles are 
cube-like rather than point-like, and they merge to a small extent. 
However, we do not just consider these rigid intrinsic shapes. 
In \cite{HKM} we showed that the two configurations are linked by a 
simple dynamical path within the Skyrme model, and we constructed a 
2-parameter manifold of four-alpha-particle configurations that interpolates 
between these most symmetric solutions, and includes bent rhomb 
(rhomboidal) configurations \cite{Bau,Brinkalpha}. All the configurations 
parametrized by this manifold have $D_2$ symmetry. They can be 
regarded as alpha particles on four alternating vertices of a cuboid with
variable shape, but fixed overall scale. Small deformations away from the 
tetrahedron, using the two parameters, correspond exactly to
deformations of a tetrahedron by the doubly-degenerate vibrational 
E-mode, so we refer to the full 2-parameter configuration space as 
the E-manifold.

In \cite{HKM} we quantized the E-manifold, taking account of
the rotational degrees of freedom. The E-manifold has a structure like 
the surface of a deformed sphere with cubic symmetry, and the 
Hamiltonian on it has both kinetic and potential terms consistent with 
this symmetry. The Hamiltonian was not derived, so our results are
phenomenological. However we found a number of states, and classified
them by energy and spin/parity. In the bigger picture, they are
the states of a rovibrational model, where there are any number of
E-phonons, but no A- or F-phonons. The paper \cite{HKM} was rather
brief, but far more detail was presented in the PhD theses of the
first two authors \cite{CJH,CK}. There, the wavefunctions of more than a dozen
E-manifold states were illustrated, and a start was made on extending
the analysis to include the A- and F-phonons. This work is developed
here, and we will clarify how the E-manifold states are interpreted in 
terms of E-phonons, which has not been done before. The Hamiltonian 
is not simply a harmonic 
oscillator in the neighbourhood of the tetrahedral solution, so the 
model gives insight into anharmonic aspects of 
vibrational E-phonons, but it is close enough to harmonic that multi-phonon 
states can easily be recognised. Being anharmonic, the model allows for 
dynamical transitions from the tetrahedron to its dual, via the
square, so it captures in a rather sophisticated way the tunnelling 
between these configurations that Dennison considered, but 
modelled rather simply \cite{Denn}. Our E-manifold Hamiltonian
predicts a significant lifting of the parity doubling, because 
the tunnelling probability is not negligible. In particular, in \cite{HKM} 
we found a good fit to the splitting between the low-lying 6.92 MeV 
$2^+$ and 8.87 MeV $2^-$ states of Oxygen-16. These are the 
lowest states with one E-phonon, and appear as degenerate in energy 
in \cite{TetVib}, for example.

In this paper, we will describe a much more complete spectrum of states 
with several E-, A- and F-phonons. We assume that one 
F-phonon has energy about 6 MeV, and one A-phonon has energy about
12 MeV. An E-phonon has a smaller energy of about 3.5 MeV, although 
in practice the E-mode excitation energies are based on E-manifold quantum 
states that take into account the anharmonicity among E-modes, and 
tunnelling. The above ordering of frequencies is novel, because most 
earlier models treated the A-mode frequency as the lowest, but this 
ordering has more than one motivation. First, in a simple tetrahedral model 
of four equal bodies connected by six equal springs, the frequencies of
the E-, F- and A-modes are in the proportions 1, $\sqrt{2}$ and
2 \cite{Wheel}. A breather state, arising from exciting the A-mode, 
therefore has rather high
energy. Second, in a related Skyrmion-inspired model of Carbon-12, a good
understanding of several excited states was achieved \cite{Raw}. There, 
the $0^+$ ground state is based on an equilateral triangular
arrangement of three alpha particles, and the 7.7 MeV Hoyle state, the 
lowest $0^+$ excited state, arises from a vibrational excitation of the 
bending mode that connects the equilateral triangle to the straight 
chain of three alpha particles. The Hoyle state then has $2^+$ 
and $4^+$ rotational excitations, in agreement with data \cite{FF}. The 
1-phonon breather excitation models the next $0^+$ state of Carbon-12 
at 9.9 MeV, whose rotational excitations would have spin/parity $2^+$, 
$3^-$ and $4^\pm$, as in the Carbon-12 ground-state band. Using this, we can 
calibrate the A-mode frequency in Oxygen-16 by using the simple spring 
model between alpha particles and comparing with Carbon-12. The ratio 
of breather frequencies for a tetrahedron with six springs and a triangle with 
three springs is $2/\sqrt{3} \simeq 1.15$, so the 9.9 MeV breather state in 
Carbon-12 implies an 11.4 MeV breather state in Oxygen-16. Our A-mode
frequency is close to this.

Oxygen-16 has a first-excited $0^+$ state at 6.05 MeV, and in previous
rovibrational models, this state was usually identified
with the 1-phonon breather state, requiring the A-mode to have
frequency 6.05 MeV. But in our model this state is instead a 2-phonon 
E-excitation, still with spin/parity $0^+$, which is why the E-mode
frequency has to be near 3 MeV. The F-mode frequency is fixed by
(uncontroversially) identifying the lowest $1^-$ state at 7.12 MeV as 
a 1-phonon F-excitation. The total energy of this state is approximately 
6 MeV arising from the F-mode frequency, plus 1 MeV from the rotational energy.

The moment of inertia calibration for our model closely follows 
Robson \cite{Rob}. The ground-state rotational band of Oxygen-16 has states
associated with a rotating tetrahedron, with spin/parities $0^+$, $3^-$, $4^+$,
$6^{\pm}, 7^-, 8^+$ and $9^{\pm}$. Because a tetrahedron has the moment of 
inertia tensor of a spherical rotor, the rotational energies are of the form 
$BJ(J+1) - C(J(J+1))^2 + D(J(J+1))^3$ where $B$ is Dennison's shorthand for 
$\frac{1}{2I}$, with $I$ the moment of inertia; to fit the $3^-$ 
state at 6.13 MeV we set $B$ to be 0.56 MeV. States beyond spin 3 have
a significant centrifugal energy correction. The leading term
$-C(J(J+1))^2$ is similar to Robson's, and we include the
term $D(J(J+1))^3$ to ensure the rotational energy continues to
increase up to spin 9 and beyond\footnote{The parameters $B$ and $C$ 
are denoted $B$ and $D_s$ in Robson's work.}. Robson chose 
$C = 3.2 \times 10^{-3}$ MeV, but we prefer a larger value 
$C = 4.5 \times 10^{-3}$ MeV combined with $D = 2.8 \times 10^{-5}$ MeV. We make 
small reductions to $B$ when fitting rotational bands of vibrationally 
excited states -- a standard correction in molecular physics
\cite{Her}, reflecting the increased moment of inertia of a vibrating state.
As in Robson's analysis \cite{Rob}, we find the ground-state rotational 
band incorporates the first-excited $6^+$ state with energy 16.27 MeV, 
whereas the lowest $6^+$ state with energy 14.82 MeV is part of the 
rotational band with one E-phonon. This band crossing is possible 
because of the difference between the moments of inertia in each band.  

The final ingredient in our model is the Coriolis correction to
the energies. This has been studied in depth by theoretical molecular chemists
since the 1930s. Herzberg gives an illuminating review \cite{Her}, but the key
original paper discussing Coriolis effects in tetrahedral molecules is by
Johnston and Dennison \cite{JD}. The Coriolis effect arises because
vibrational motion can carry an internal angular momentum, and this
influences the total rotational energy. Vibrational excitations
only involving A- and E-phonons have no Coriolis energy correction, but the
F-band states -- the states in the rotational band with just one 
F-phonon -- do. A similar correction occurs when an F-phonon is
combined with any number of A- and E-phonons. A rather different 
Coriolis correction occurs in the rotational band with two F-phonons. 
These corrections are reviewed in Section 3. Our most important observation is 
that the zeta factor, occurring in the Coriolis-corrected rotational 
energies, is not $\zeta = -0.5$. This is the value for four 
point-like alpha particles. (It is the value that emerges
from an analysis of methane, when the central carbon atom is decoupled.) 
Dennison \cite{Denn} assumed that $\zeta = -0.5$ in the Oxygen-16 nucleus, 
and this assumption has been adopted by others \cite{Rob,TetVib}. However,
for extended alpha particle structures, as in the Skyrmion picture, where
the alpha particles are partly merged and are not vibrating and
rotating in the same way as
hydrogen atoms do in methane, one can contemplate a different zeta
value. Robson mentions this possibility in \cite{Rob1}. Values of 
$\zeta$ in the range $-1$ to $1$ are common for molecules with 
various geometries, and we find that a 
best fit occurs for $\zeta$ close to 0.2. The main reason for
our choice is to fit the splitting between the 11.08 MeV $3^+$ state and the
11.60 MeV $3^-$ state in Oxygen-16. These are both modelled as F-band states,
and the splitting is largely due to the Coriolis effect.

With these assumptions for our model and its parameters, we have 
calculated a rovibrational energy spectrum for the Oxygen-16 nucleus where
essentially all states up to 20 MeV excitation energy are fitted moderately
well. There are just over 60 such states with isospin 0 in the experimental
tables \cite{O16Exp,ENSDF}, although there are uncertainties in the
spin/parity and isospin assignments for several of the higher-energy 
states. The vibrational states we need to consider include up 
to four E-phonons, or two F-phonons, but at most one A-phonon. One new
$4^-$ state is predicted below 15 MeV. There are no confirmed $6^-$ 
states in the tables, but our model predicts a few of these above 
17 MeV. It also predicts several higher-spin states between 20 MeV 
and 30 MeV. Our model relies on seven basic parameters --
the three vibrational frequencies $\omega_E$, $\omega_F$ and 
$\omega_A$, and the four rotational parameters $B, C, D$ and $\zeta$. In 
addition, there are some physically motivated adjustments to $B$ that are 
fitted, depending on the vibrational state. Actually, this parameter 
count is a simplification because the E-vibrational energies are 
obtained from the quantization of the E-manifold, which has its own 
model, rather than by simply counting E-phonons.

\section{E-manifold states}

Here we consider the states that arise from quantization of
the E-manifold. They include states in the ground-state rotational 
band, and in all the rotational bands associated with E-phonon vibrational 
excitations. The model for the E-manifold was introduced in
ref. \cite{HKM}. It is reviewed here, and we
present the wavefunctions of the states, their energies and spin/parities in more detail than in \cite{HKM}. Most of these
results appeared previously in \cite{CJH,CK}, but have been extended
and updated here. We also clarify, for the first time, how the
E-manifold states can be classified in terms of E-phonon counting.

\begin{figure}
	\begin{center}
		\includegraphics[width=0.80\textwidth]{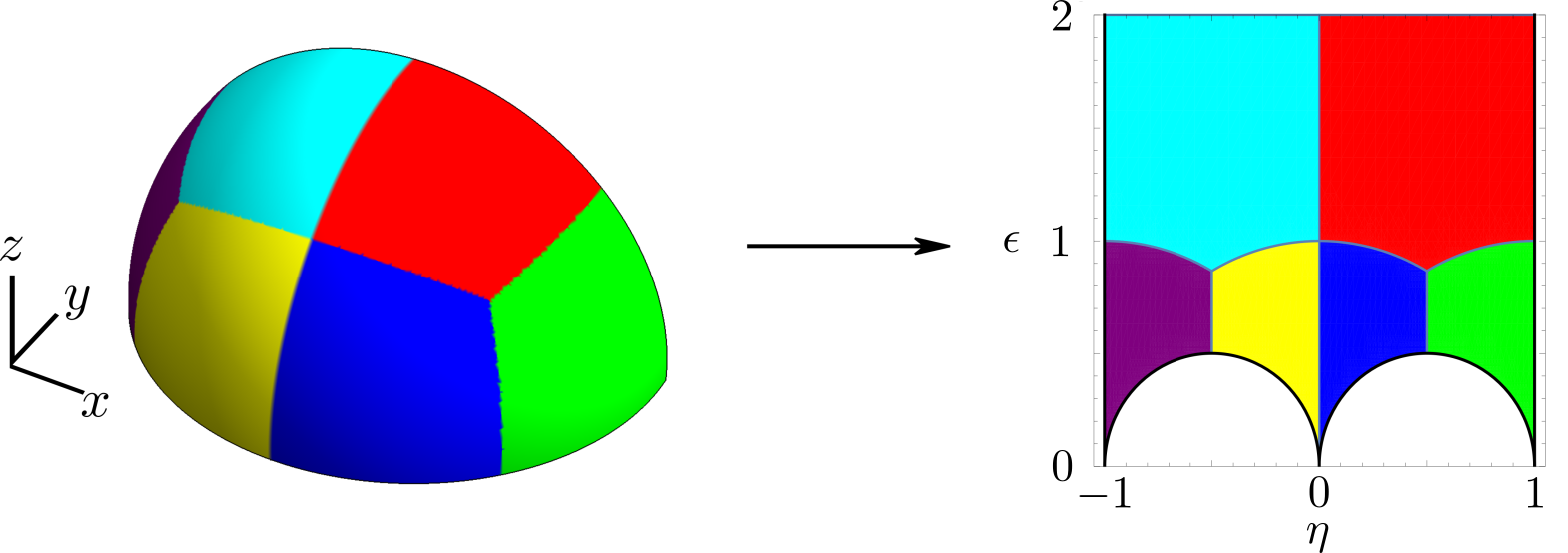} 
	\end{center}
\caption{The correspondence between the sphere and complex 
plane for one quarter of the E-manifold. Coloured regions are 
mapped to one another.} \label{mapping}
	
\end{figure}

\begin{figure}
	\begin{center}
		\includegraphics[width=0.40\textwidth]{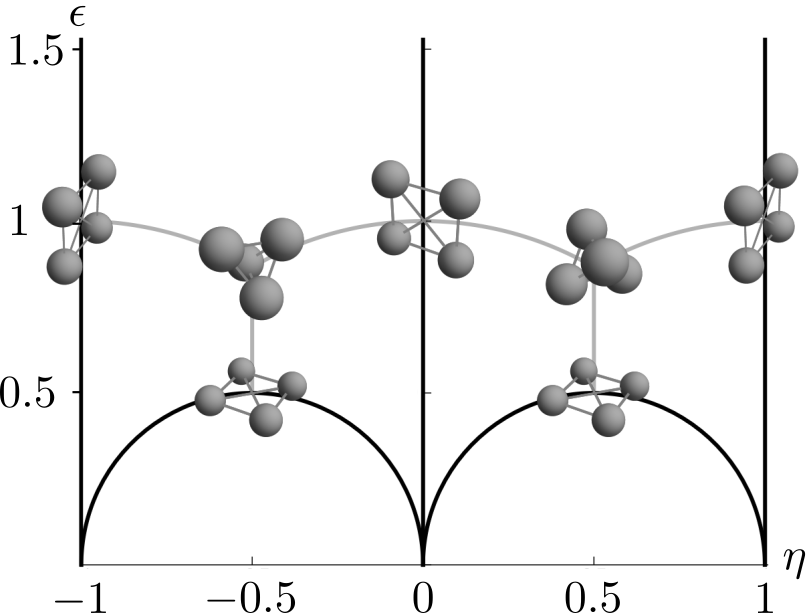} 
	\end{center}
\caption{The locations of some symmetric configurations on the complex 
plane. The balls represent alpha particles.} \label{configs}
	
\end{figure}

The E-manifold is a model for four point alpha particles arranged in
configurations with $D_{2}$ symmetry. The centre of mass is at the
origin, and the $D_{2}$ symmetry is with respect to standard,
Cartesian body-fixed axes. Let the Cartesian coordinates of one 
alpha particle be $(x,y,z)$; the other three are then at 
$(x,-y,-z)$, $(-x,y,-z)$ and $(-x,-y,z)$, so the
four particles are at alternating vertices of a cuboid. The scale size
is fixed for each cuboid shape using a potential that 
disfavours the alpha particles from being too close together or too 
far apart. The E-manifold is therefore topologically a 2-sphere,
parametrised by the direction of $(x,y,z)$. To 
visualise the E-manifold, we project a quarter of the 2-sphere onto a
region of the complex plane, as shown in 
Fig. \ref{mapping}. Each point on the plane
corresponds to a configuration of the four alpha particles, and the
positions and orientations of the tetrahedron and square 
configurations, on the complex plane, are shown in Fig. \ref{configs}.

The potential on the E-manifold has a minimum at the
tetrahedral configuration, $(x,y,z) = (1,1,1)$ modulo scale, and there is
a saddle point at the square configuration, $(x,y,z) = (1,1,0)$
modulo scale. The dynamics on the E-manifold is also arranged to force
wavefunctions to approach zero as $(x,y,z)$ approaches $(1,0,0)$
modulo scale, where the alpha particles form two pairs with a large 
separation between them.

The action of $D_2$ ($\pi$-rotations about the Cartesian axes) permutes
the alpha particles, and since these are bosons, wavefunctions
must be $D_2$-invariant. The full symmetry group of the E-manifold,
and of the dynamics on it, is the
cubic group $O_h = O \times \ZZ$, with 48 elements. The non-trivial
element of $\ZZ$ is inversion, $(x,y,z) \to (-x,-y,-z)$. The quotient
of $O$ by its normal subgroup $D_2$ gives the permutation group $S_3$,
which permutes the (unoriented) Cartesian axes. As a
result, allowed wavefunctions on the E-manifold are classified by a
representation of $S_3$, and also a sign depending on how $\ZZ$ acts.

$S_3$ has three irreducible representations (irreps) -- the 1-dimensional 
trivial irrep $T$, the 1-dimensional sign irrep $S$, and the
2-dimensional standard irrep $St$. Wavefunctions on the E-manifold are
therefore labelled by species $T^+$, $S^+$, $St^+$ or $T^-$, $S^-$, $St^-$ with
an additional subscript $n$ to denote the number of phonons in the
state (which is described carefully below). The
wavefunctions of species $T$ and $S$ are denoted by $\psi$, but those 
of species $St$ have a 2-fold degeneracy, and the
two linearly independent wavefunctions are denoted $u$ and $v$.

\begin{figure}

	\begin{subfigure}{0.23\textwidth}
		\centering
		\includegraphics[width=\textwidth]{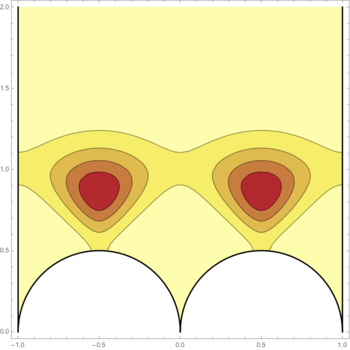} 
		\caption*{$\psi^{+}_{T0},\, E_\text{vib} = 0$}
	\end{subfigure}
	\begin{subfigure}{0.23\textwidth}
		\includegraphics[width=\textwidth]{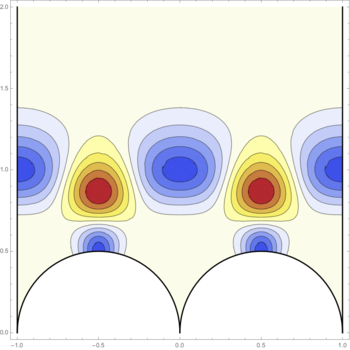} 
		\caption*{$\psi^{+}_{T2},\, E_\text{vib} = 6.05$}
	\end{subfigure}
	\begin{subfigure}{0.23\textwidth}
		\includegraphics[width=\textwidth]{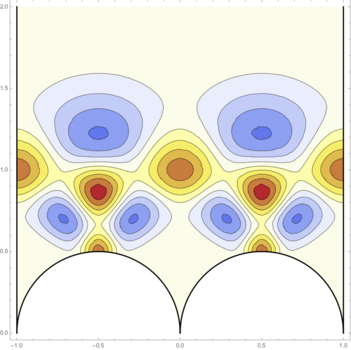} 
		\caption*{$\psi^{+}_{T3},\, E_\text{vib} = 14.89$}
	\end{subfigure}
	\begin{subfigure}{0.23\textwidth}
		\includegraphics[width=\textwidth]{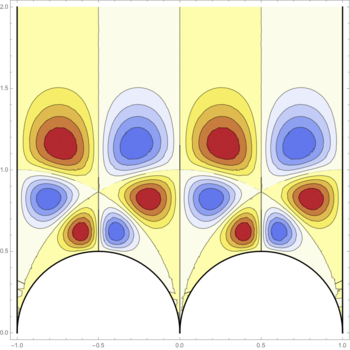} 
		\caption*{$\psi^{-}_{T3},\, E_\text{vib} = 16.35$}
	\end{subfigure}

	\begin{subfigure}{0.23\textwidth}
		\includegraphics[width=\textwidth]{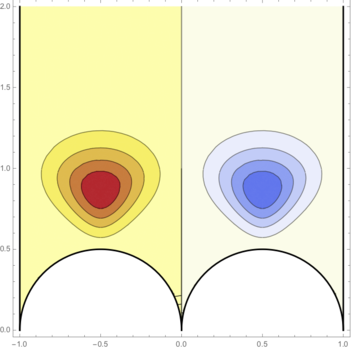} 
		\caption*{$\psi^{-}_{S0},\, E_\text{vib} = 0.18$}
	\end{subfigure}
	\begin{subfigure}{0.23\textwidth}
		\includegraphics[width=\textwidth]{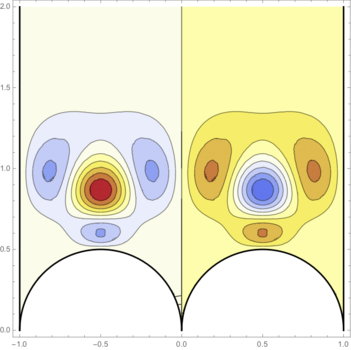} 
		\caption*{$\psi^{-}_{S2},\, E_\text{vib} = 10.67$}
	\end{subfigure}
	\begin{subfigure}{0.23\textwidth}
		\includegraphics[width=\textwidth]{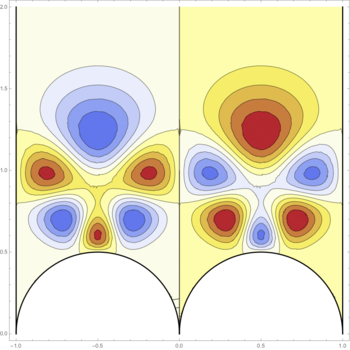} 
		\caption*{$\psi^{-}_{S3},\, E_\text{vib} = 16.68$}
	\end{subfigure}
	\begin{subfigure}{0.23\textwidth}
		\includegraphics[width=\textwidth]{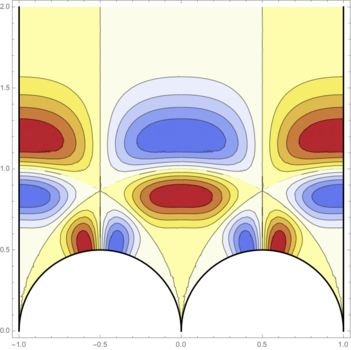} 
		\caption*{$\psi^{+}_{S3},\, E_\text{vib} = 12.57$}
	\end{subfigure}

	\begin{subfigure}{0.32\textwidth}
		\includegraphics[width=0.48\textwidth]{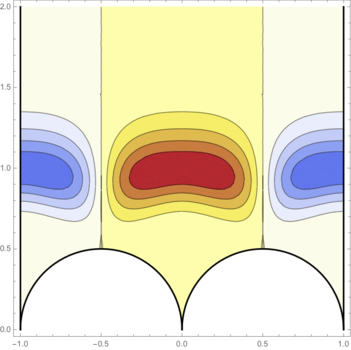} 
		\includegraphics[width=0.48\textwidth]{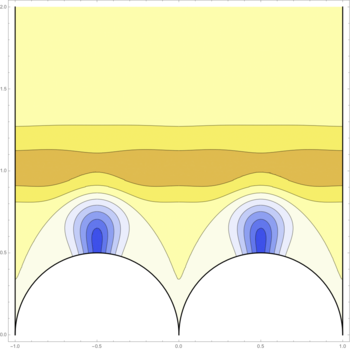} 
		\caption*{\centering$u_1^+-v_1^+$ and $u_1^++v_1^+, \linebreak E_\text{vib} = 3.45$}
	\end{subfigure}
	\begin{subfigure}{0.32\textwidth}
		\includegraphics[width=0.48\textwidth]{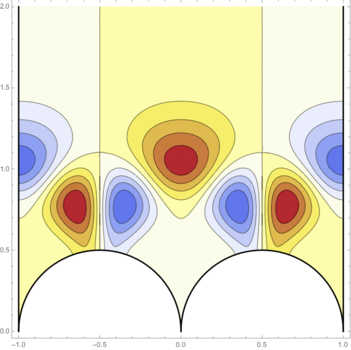} 
		\includegraphics[width=0.48\textwidth]{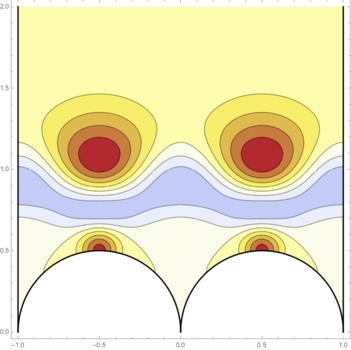} 
		\caption*{\centering$u_2^+-v_2^+$ and $u_2^++v_2^+, \linebreak E_\text{vib} = 8.72$}
	\end{subfigure}
	\begin{subfigure}{0.32\textwidth}
		\includegraphics[width=0.48\textwidth]{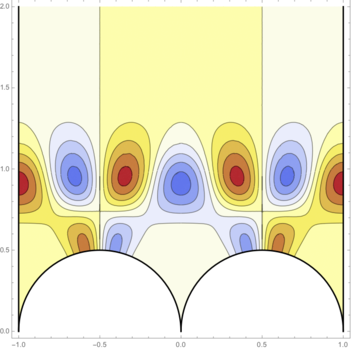} 
		\includegraphics[width=0.48\textwidth]{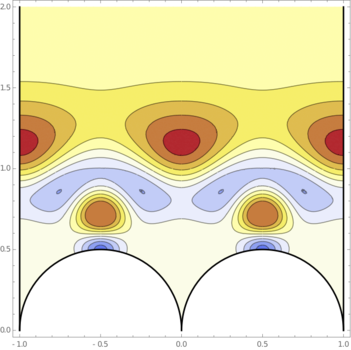} 
		\caption*{\centering $u_3^+-v_3^+$ and $u_3^++v_3^+, \linebreak E_\text{vib} = 12.22$}
	\end{subfigure}
	
	\begin{subfigure}{0.32\textwidth}
		\includegraphics[width=0.48\textwidth]{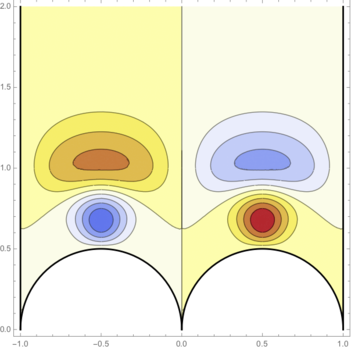} 
		\includegraphics[width=0.48\textwidth]{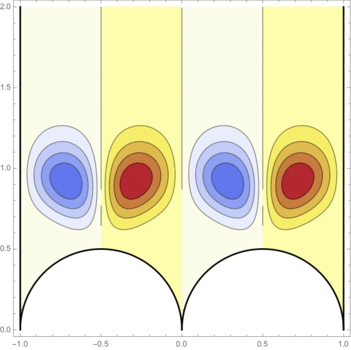} 
		\caption*{\centering$u_1^--v_1^-$ and $u_1^-+v_1^-, \linebreak E_\text{vib} = 5.27$}
	\end{subfigure}
	\begin{subfigure}{0.32\textwidth}
		\includegraphics[width=0.48\textwidth]{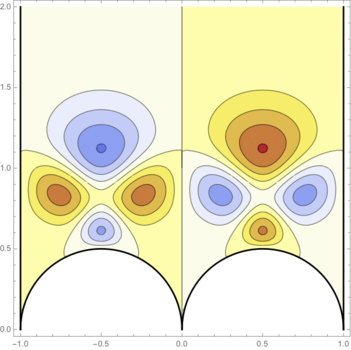} 
		\includegraphics[width=0.48\textwidth]{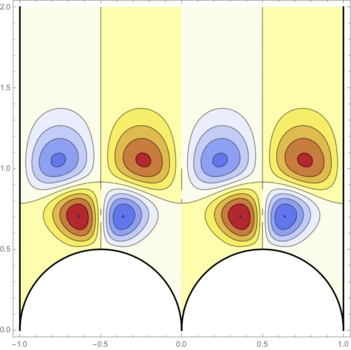} 
		\caption*{\centering$u_2^--v_2^-$ and $u_2^-+v_2^-, \linebreak E_\text{vib} = 11.05$}
	\end{subfigure}
	\begin{subfigure}{0.32\textwidth}
		\includegraphics[width=0.48\textwidth]{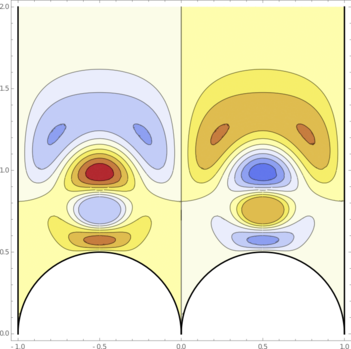} 
		\includegraphics[width=0.48\textwidth]{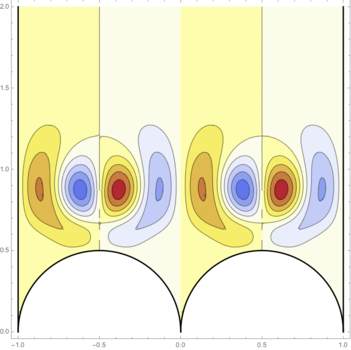} 
		\caption*{\centering$u_3^+-v_3^-$ and $u_3^-+v_3^-, \linebreak E_\text{vib} = 16.92$}
	\end{subfigure}

\caption{The vibrational wavefunctions on the E-manifold. The
plots show the wavefunction contours from $-1$ (blue) to $+1$ (red). 
The wavefunctions in the first row transform as the trivial irrep $T$, 
while those in the second row transform as the sign irrep $S$. The third 
(fourth) row transform as the standard irrep $St$ with positive 
(negative) parity. The wavefunctions and axes are scaled for clarity. 
Energies are in MeV.} \label{wavefunctions}

\end{figure}

Our Hamiltonian combines a kinetic term on the E-manifold
(based on a hyperbolic metric on the 6-punctured sphere), and a
potential that disfavours the alpha particles splitting into
two pairs \cite{HKM}.  
We have investigated numerically all the low-energy wavefunctions on
the E-manifold. They are illustrated in Fig. \ref{wavefunctions}, 
alongside their energies $E_{\rm vib}$. The
two lowest states $\psi_{T0}^+$ and $\psi_{S0}^-$ have wavefunctions
concentrated around the potential minima at the tetrahedron $(1,1,1)$
and its dual $(-1,-1,-1)$. Tunnelling between these is via the square 
saddle point. Wavefunctions of the species $S$ are constrained to 
vanish at the square while the species $T$ wavefunctions have no such
constraint. When the tunnelling amplitude is large, there is
a relatively large energy gap between states that
would otherwise be degenerate parity doubles. The lowest-energy wavefunctions,
$\psi_{T0}^+$ and $\psi_{S0}^-$, have a small tunnelling amplitude, so
their energy gap is small. 
Rotational excitations of these states jointly form the ground-state band. 
Higher-energy wavefunctions of species $\psi_T^+$ and $\psi_S^-$ have 
a larger energy gap,  because tunnelling is easier. The
state $\psi_{T2}^+$ will be identified with the low-lying $0^+$ state
at $6.05$ MeV. It is concentrated around both the tetrahedron and the
square configurations. Physically, this state should be thought of as
an admixture of a tetrahedron and a square. The lowest-energy states
of species $St$ are a positive-parity pair $u_1^+ + v_1^+$ and 
$u_1^+ - v_1^+$, and a negative-parity pair $u_1^- + v_1^-$ and 
$u_1^- - v_1^-$. The positive parity states are concentrated around the square 
configurations while the negative parity states are concentrated around the (prolate and oblate) bent rhombs. The rotational bands arising from these, with spin/parity 
$2^\pm, 4^\pm, 5^\pm, ...$ can be interpreted as rotational excitations of 
the squares and rhombs.

A rather different classification is possible, and important
here. Within $O_h$ there is the 24-element subgroup $T_d$. This is the subgroup
that preserves the configuration of four alpha particles at the
vertices of a regular tetrahedron, where $(x,y,z) = (1,1,1)$. $T_d$
acts nontrivially, but linearly, on small deformations of a configuration
away from a regular tetrahedron. The group $T_d$ has five irreps, the
trivial and nontrivial 1-dimensional irreps $A_1$ and $A_2$, the
2-dimensional irrep $E$, and the two 3-dimensional irreps $F_1$ and
$F_2$ \cite{LL}. (The 3-dimensional irreps are often written as $T_1$ 
and $T_2$ \cite{Cot}.)

Vibrational modes away from the tetrahedron in the direction of the
E-manifold are 2-fold degenerate, and transform under the irrep $E$ of
$T_d$. Both the states $\psi_{T0}^+$ and $\psi_{S0}^-$ are invariant under
$T_d$, and are interpreted as states with no E-phonons, but the pair
of states $u_1^+ + v_1^+$ and $u_1^+ - v_1^+$ are interpreted as degenerate
1-phonon states transforming under the irrep $E$, as are 
the pair $u_1^- + v_1^-$ and $u_1^- - v_1^-$. 
As $T_d$ is a subgroup of $O_h$, each E-manifold state
classified by an irrep of $O_h$ is also classified by a $T_d$
irrep. The states of species $T^+$ and $S^-$ are in the $A_1$ irrep of
$T_d$, the states of species $T^-$ and $S^+$ are in the $A_2$ irrep,
and the states of species $St^+$ and $St^-$ are in the $E$ irrep. No
E-manifold states are classified by $F_1$ or $F_2$. The $T_d$
classification misses the $\ZZ$ label. This can be read off from the
behaviour of the wavefunction under reflection in the vertical 
line splitting in half each subfigure of Fig. \ref{wavefunctions}. 
If the wavefunction changes sign, then it is $\ZZ$-odd.

It is known how multi-phonon states of a tetrahedron vibrating in the
direction of the E-manifold transform under $T_d$. The phonons are 
bosonic, so one needs the decompositions of the symmetrised $n$th powers of 
the irrep $E$. For these, we should use the notation $E^n_{\rm symm}$, 
but shorten this to $E^n$. These decompositions are \cite{Her}, p.127,
\bea
E^0 &=& A_1 \,, \\
E^1 &=& E \,, \\
E^2 &=& A_1 \oplus E \,, \\
E^3 &=& A_1 \oplus A_2 \oplus E \,, \\
E^4 &=& A_1 \oplus 2E \,.
\eea
The dimension of $E^n$ is $n+1$, as expected for $n$-phonon states of
a 2-dimensional harmonic oscillator. Combining the above algebraic information 
with the energy estimate that an $n$-phonon state has approximately 
$n$ times the energy of a 1-phonon state, and inspecting the 
shape of the wavefunctions in Fig. \ref{wavefunctions} close to the
tetrahedral configuration, one can classify these wavefunctions
as follows. The wavefunctions $\psi_{T0}^+$, $\psi_{S0}^-$  are 
0-phonon states, $u^+_1 \pm v^+_1$, $u^-_1 \pm v^-_1$ are 1-phonon states, 
$\psi^+_{T2}, \psi^-_{S2}$ are 2-phonon $A_1$ states and 
$u^+_2 \pm v^+_2$, $u^-_2 \pm v^-_2$ are 2-phonon $E$ states. The wavefunctions 
$\psi^+_{T3}, \psi^+_{S3}$ are 3-phonon $A_1$ states, 
$\psi^-_{T3}, \psi^-_{S3}$ are 3-phonon $A_2$ states and 
$u^+_3 \pm v^+_3$, $u^-_3 \pm v^-_3$ 
are 3-phonon $E$ states.  In all cases there is a pair of states 
distinguished by the $\ZZ$ label. This classification is verified 
by looking at the nodes of the wavefunctions, and comparing with 
harmonic oscillator states near the tetrahedral point expressed in 
(plane) polar coordinates. For example, 1-phonon states have one node, 
and 2-phonon $A_1$ states have a radial node. 3-phonon states have either 
six nodes in the angular direction, or a radial node and fewer 
than six nodes in the angular direction. 

Classifying the E-manifold states in terms of E-phonons helps us
compare our list of possible states, and their
spin/parities, with the lists created by others. However, for the energy 
of an $n$-phonon state we use the E-manifold energies $E_{\rm vib}$ 
shown in Fig. \ref{wavefunctions}, rather than the harmonic estimate
$n\omega_E$. These are noticeably different, showing the importance of 
tunnelling through the square configuration. 

Each E-manifold state has a band of rotational excitations. 
The allowed spin/parities $J^P$ are those for which the 
decomposition of the $O(3)$ representation with spin/parity $J^P$ 
contains the irrep of $O_h$ classifying the E-manifold state (equivalently,
a $T_d$ irrep and a $\ZZ$ sign). These are essentially the same as 
the allowed spin/parities for vibrational states classified by 
$T_d$ irreps (ignoring the $\ZZ$ sign), which are
\cite{Her}, p.450,
\bea
A_1 & \longrightarrow & 0^+, 3^-, 4^+, 6^\pm, 7^-, 8^+, 
9^\pm, \dots \,, \label{A1list} \\  
A_2 & \longrightarrow & 0^-, 3^+, 4^-, 6^\pm, 7^+, 8^-, 
9^\pm, \dots \,, \label{A2list} \\
E & \longrightarrow & 2^\pm, 4^\pm, 5^\pm, 6^\pm, 7^\pm, 8^\pm, 8^\pm,
9^\pm, \dots \,, \label{Elist} \\
F_1 & \longrightarrow & 1^+, 2^-, 3^\pm, 4^\pm, 5^\pm, 5^+, 6^\pm,
6^-, 7^\pm, 7^\pm, 8^\pm, 8^\pm, 9^\pm, 9^\pm, 9^+, \dots \,, \label{F1list} \\
F_2 & \longrightarrow & 1^-, 2^+, 3^\pm, 4^\pm, 5^\pm, 5^-, 6^\pm,
6^+,  7^\pm, 7^\pm, 8^\pm, 8^\pm, 9^\pm, 9^\pm, 9^-, \dots \,. \label{F2list}
\eea
E-manifold states only occur in the irreps $A_1$, $A_2$ and $E$. The
extra $\ZZ$ label for E-manifold states means that we can separate
the positive and negative parity states, and assign them different
vibrational energies. So, for example, the wavefunctions 
$\psi_{T0}^+, \psi_{T2}^+, \psi_{T3}^+$ all allow positive parity $A_1$ 
states $0^+, 4^+, 6^+, \dots$, whereas the wavefunctions 
$\psi_{S0}^-, \psi_{S2}^-, \psi_{S3}^-$ allow negative parity $A_1$ states 
$3^-, 6^-, 7^-, \dots$. The wavefunction $\psi_{T3}^-$ allows the unnatural 
parity $A_2$ state $0^-$, and $\psi_{S3}^+$ the unnatural parity 
$A_2$ state $3^+$. The wavefunctions $u^\pm_n \pm v^\pm_n$ allow $E$
states $2^\pm, 4^\pm, 5^\pm, 6^\pm, \dots$.

\section{Rovibrational Energies}

We now consider the rotational excitations of vibrational states, 
where each vibrational state is classified by the
number of its A-, E- and F-phonons. We discuss all states
with total energy up to 20 MeV, and states with spin 6 and higher
up to 30 MeV. This means there are at most four E-phonons, two 
F-phonons, or one A-phonon. Combined vibrational states are also 
allowed, and include those combining one F- or A-phonon with one or 
two E-phonons. More precisely, the vibrational states are combinations 
of E-manifold wavefunctions with harmonic oscillator states
for $n_A$ A-phonons and $n_F$ F-phonons, so the vibrational energy is 
the sum of the E-manifold energy $E_{\rm vib}$ and a contribution $n_A \omega_A 
+ n_F \omega_F$. The E-manifold energy takes into account tunnelling 
between the tetrahedron and its dual, so we do not need to add an explicit
tunnelling energy as in \cite{Rob}.

The rotational energy in a spin $J$ state, including centrifugal 
corrections, is taken to be 
\be
E_{\rm rot} = BJ(J+1) - C(J(J+1))^2 + D(J(J+1))^3 \,. 
\label{Erot}
\ee 
In the ground-state band with no phonons, we set $B = 0.56$. For all 
E-manifold states having at least one E-phonon 
we set $B = 0.45$. This can be interpreted physically: the 0-phonon 
wavefunctions are concentrated at the tetrahedral configuration while the 
1-phonon E-wavefunctions are concentrated at the square. The square has a 
larger moment of inertia and hence a smaller $B$. The same value $B=0.45$ is
used for states with one F-phonon or one A-phonon, and no E-phonons. 
For states with two F-phonons, combined E- and F-phonons, or combined 
E- and A-phonons, we set $B=0.4$. This steady decrease of $B$ as the 
number of phonons increases resembles the pattern used in molecular 
physics \cite{Her}. For the parameters $C$ and $D$ we choose the
constant values $C = 4.5 \times 10^{-3}$ and $D = 2.8 \times
10^{-5}$. This ensures that the rotational energy increases
approximately linearly with $J$ between $J=3$ and $J=8$. 
The formula (\ref{Erot}) is a simplified version of
one proposed by Sood \cite{Soo,Wol}.  

This is all we need to calculate the energy of states
with no F-phonons. Mostly, these are multiple E-phonon states with 
no A-phonons, but some have a single A-phonon excitation 
transforming under the irrep $A_1$. Since the $A_1$ irrep is trivial, the 
vibrational species $A_1 \otimes E^n$ allows for the same spins and 
parities as the $E^n$ species.

There is a Coriolis energy correction for the F-band states, i.e. states 
with a single F-phonon and no A- or E-phonons. Each F-band state has
an underlying E-manifold state with definite spin/parity situated in 
the ground-state rotational band, and based on either $\psi_{T0}^+$ or 
$\psi_{S0}^-$. This is combined using the usual Clebsch--Gordon angular
momentum rules with a 1-phonon F-mode excitation that carries
internal angular momentum 1 and (being triply-degenerate) transforms 
as the $F_2$ irrep of the $T_d$ group. The states from the 
ground-state rotational band have spin/parities 
$0^+, 3^-, 4^+, 6^\pm, \dots$, whose rotational angular momentum is denoted by 
$\bR$; the intrinsic angular momentum of the F-phonon is denoted by 
$\bl$. $\bR$ and $\bl$ commute, and the total angular momentum is
\be
\bJ = \bR + \bl \,.
\label{R+l}
\ee
The parity of the combined state is the opposite of the parity of the
underlying rotational state, because the F-phonon has negative parity.
The allowed combined states therefore have spin/parities (up to spin 9)
\be
1^-_0, 2^+_3, 3^+_3, 4^+_3, 3^-_4, 4^-_4, 5^-_4, 5^\pm_6, 6^\pm_6,
7^\pm_6, 6^+_7, 7^+_7, 8^+_7, 7^-_8, 8^-_8, 9^-_8, 8^\pm_9, 9^\pm_9,
9^\pm_{10} \,,
\label{Fband}
\ee
where the usual spin/parity label $J^P$ is supplemented by a
subscript $R$ to denote the underlying rotational angular
momentum. $R$ has one of the values $J+1$, $J$ or $J-1$. Note that the 
$J^P$ values occurring here are exactly the same as those one finds when 
considering an F-phonon as transforming under the $F_2$ irrep of $T_d$
(see the list (\ref{F2list})).

The total rotational energy, including the Coriolis and centrifugal 
corrections, arises from the Hamiltonian \cite{JD,Her} 
\be
H_{\rm rot} = B(\bJ - \zeta \bl)^2 - C(J(J+1))^2 + D(J(J+1))^3 \,.
\label{Hrot}
\ee
Expanding out, this is
\be
H_{\rm rot} = BJ(J+1) - 2B\zeta \bJ \cdot \bl + 2B\zeta^2 -
C(J(J+1))^2 + D(J(J+1))^3 \,,
\ee
where we have set $l(l+1) = 2$ for internal angular momentum $l=1$.
(This expansion is valid even though $\bJ$ and $\bl$ do not commute, because the
component pairs $\bJ_i$ and $\bl_i$ do commute.) By squaring eq. (\ref{R+l}) 
we find $2\bJ \cdot \bl = J(J+1) - R(R+1) + 2$, so the energy
eigenvalues of $H_{\rm rot}$ are
\bea
E_{\rm rot} &=& BJ(J+1) - 2B\zeta(1-\zeta) - C(J(J+1))^2 + D(J(J+1))^3
\nonumber \\
&& \quad\quad + \, 2B\zeta 
\begin{cases}
J+1 & \text{if} \quad R = J+1 \,, \\ 
0 & \text{if} \quad R = J \,, \\
-J & \text{if} \quad R = J-1 \,.
\end{cases}
\label{ErotF}
\eea 

We now need to fix a calibration for $\zeta$. As mentioned earlier, we will
not make the standard choice, $\zeta = -0.5$. Instead, we fit $\zeta$
using the energies of
the $3^+_3$ and $3^-_4$ states in the F-band. The lowest F-band
state should clearly be identified with the experimental $1^-$ state 
at 7.12 MeV, and spin 3 states are 4 MeV to 5 MeV above this. There is 
just one experimentally confirmed $3^+$ state, at 11.08 MeV, and we
identify this with the $3^+_3$ state in the F-band. We identify
the $3^-_4$ state in the F-band with the experimental $3^-$ state at 
11.60 MeV (this is the first-excited $3^-$ state, as the lowest such 
state at 6.13 MeV is in the ground-state band, with no phonons).  
In our model, there are two sources for the 0.52 MeV energy 
splitting $E(3^-) - E(3^+)$. First, the underlying E-manifold states 
have an energy splitting of $-0.18$ MeV, because the $R=4$ state with 
positive parity has underlying state $\psi_{T0}^+$ and the $R=3$ state 
with negative parity has underlying state $\psi_{S0}^-$. The 
additional 0.70 MeV is from the Coriolis splitting between $J=3$ states 
with $R=4$ and $R=3$. This requires the calibration $2B\zeta = 0.175$, 
and as $2B = 0.9$ for the F-band, $\zeta = 0.194$. We will see below
that this positive value for $\zeta$ gives reasonable 
energy splittings for several other states, up to spin/parity $5^\pm$.

With $\zeta$ fixed, we note that the term
$-2B\zeta(1-\zeta)$ in eq. \eqref{ErotF} is constant for the entire F-band and has 
value $-0.14$ MeV. This could be absorbed into the F-band phonon 
frequency, but we do not do this, as different constants occur in 
other bands, in particular the 2-phonon ${\rm F}^2$-band.

The rotational energy formula \eqref{ErotF} extends to combined bands. It is 
known from molecular physics that $\zeta$ is unchanged for all bands of species
$E^n \otimes F_2$. From the tetrahedral representation theory, it is 
known that the ${\rm E} \times {\rm F}$ band, with one E- and one F-phonon, 
transforms as $F_1 \oplus F_2$, and therefore the allowed
spin/parities are parity doubles of the F-band states listed in \eqref{Fband}. 
Again, there is an equivalent interpretation of these
states. They are combinations of an underlying rotational state with one
E-phonon, having spin/parity $2^\pm, 4^\pm, 5^\pm, \dots$, and an
F-phonon with internal spin/parity $1^-$. For example, 
in this band there are low-spin states with $J^P = 1^\pm, 2^\pm, 3^\pm$ all of 
which have $R=2$, and further $3^\pm$ states with $R=4$.    

In the ${\rm E}^2 \times {\rm F}$ band, the calculation is similar. We recall
that $E^2 = A_1 \oplus E$, so states with two E-phonons have
underlying rotational states in the two lists (\ref{A1list}) and
(\ref{Elist}). The angular momentum $R$ of the underlying state combines
with the $1^-$ of the F-phonon to give the total $J^P$. As the
vibrational energy is quite high, we need only consider states with
spins up to $J=3$ in this band. We are less confident about our energies
for states in the ${\rm E} \times {\rm F}$ and 
${\rm E}^2 \times {\rm F}$ bands, because the relatively large
amplitude of E-phonon(s) has an anharmonic effect on the F-phonon.

Finally, in the ${\rm F}^2$-band, things are a little different. States here
have no E-phonons, so the underlying rotational states are those of
the ground-state band with spin/parities $0^+, 3^-, 4^+, \dots$. As a 
single F-phonon has spin/parity $1^-$, two F-phonons have internal spin/parity 
$0^+$ or $2^+$. Therefore $l=0$ or $l=2$ in the ${\rm F}^2$-band. 
There are six states here, and this is consistent with the 
$T_d$ decomposition $(F_2 \otimes F_2)_{\rm symm} = A_1 \oplus E \oplus F_2$. 
For the $l=0$ state there is no Coriolis effect,
and the rotational energy (without the centrifugal corrections) is 
simply $BJ(J+1)$. Here $B=0.4$, the 
reduced value associated with having two F-phonons. For the 
$l=2$ states, the rotational contribution to the energy is
calculated starting from $H_{\rm rot}$, the Hamiltonian in 
eq. (\ref{Hrot}), with $B=0.4$ and $\zeta = -0.194$. The reversal of 
the sign of $\zeta$ follows the calculations in molecular physics 
\cite{JD}, although it may not be justified here. Expanding out 
$H_{\rm rot}$ we find rotational energies
\bea
E_{\rm rot} &=& BJ(J+1) - 6B\zeta(1-\zeta) - C(J(J+1))^2 + D(J(J+1))^3 
\nonumber \\ 
&& \quad\quad + \, 2B\zeta 
\begin{cases}
2J+3 & \text{if} \quad R = J+2 \,, \\
J+1 & \text{if} \quad R = J+1 \,, \\ 
0 & \text{if} \quad R = J \,, \\
-J & \text{if} \quad R = J-1 \,, \\
-2J+1 & \text{if} \quad R = J-2 \,,
\end{cases}
\label{ErotF2}
\eea 
where $R = 0,3,4, \dots$. States in the ${\rm F}^2$-band with energy 
below 20 MeV have spins no greater than $J=4$. The parity of each 
state is the same as that of the underlying state, $0^+$, $3^-$ or $4^+$.

\section{The Oxygen-16 Energy Spectrum}

\begin{figure}
	\begin{center}
		\includegraphics[width=1\textwidth]{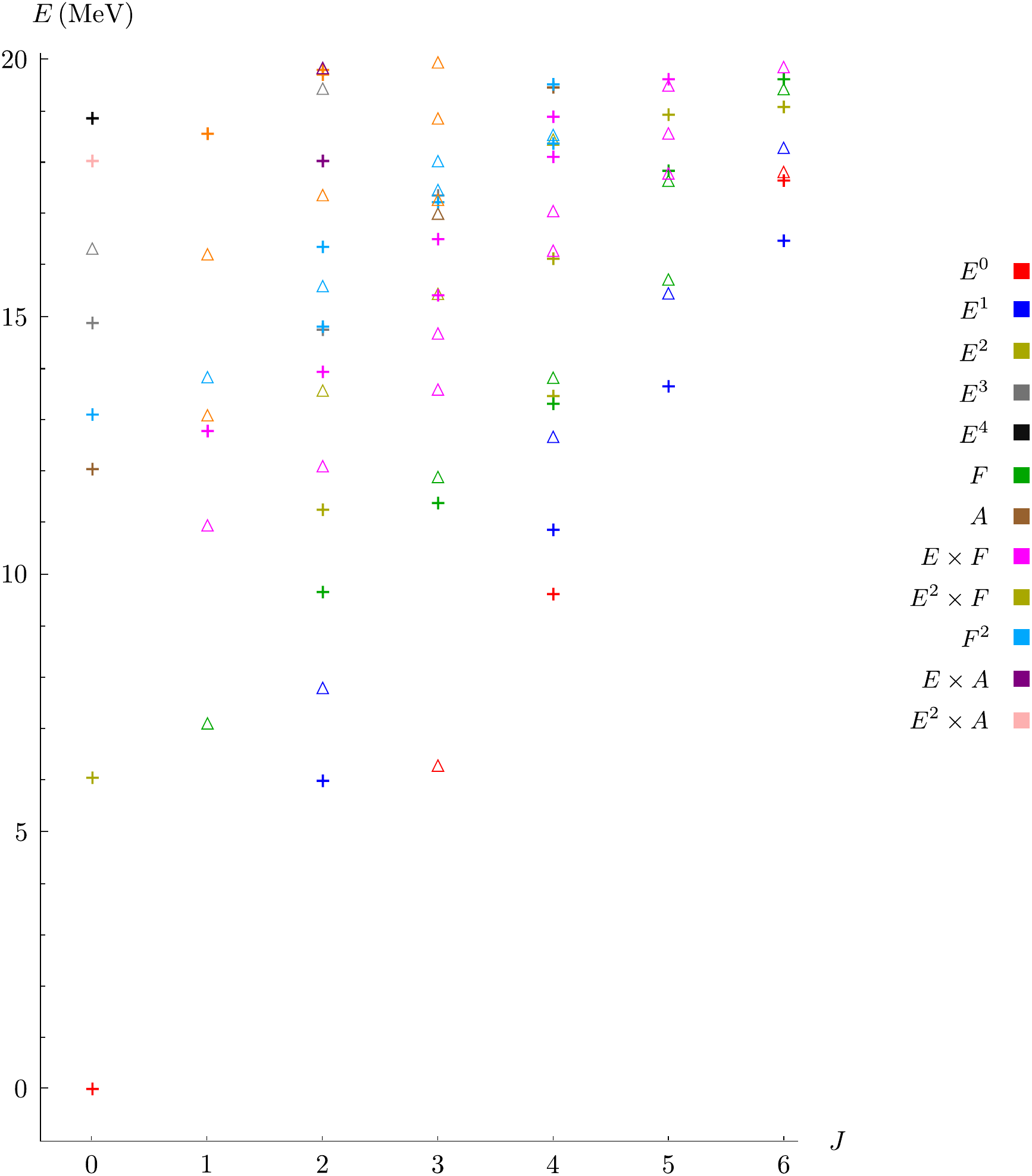} 
	\end{center}
	\caption{Theoretical energy spectrum of our model. Each
          rotational band is coloured differently. Positive (negative) 
          parity states are displayed as pluses (triangles). } \label{spectrum}	
\end{figure}

\begin{figure}
	\begin{center}
		\includegraphics[width=0.8\textwidth]{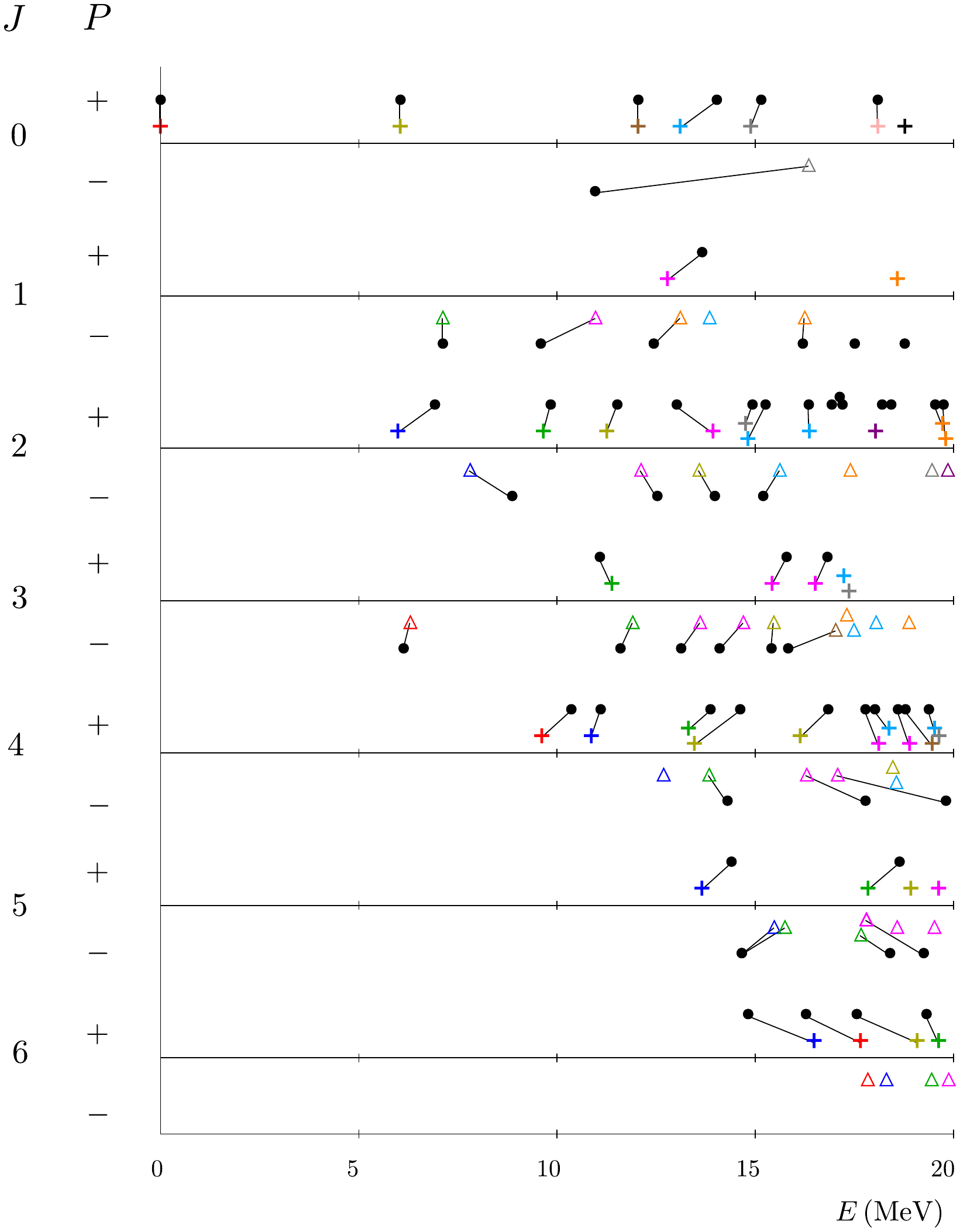} 
	\end{center}
	\caption{Comparison between theoretical and experimental
          energies. Positive (negative) parity states of our model are
          displayed as pluses (triangles), and coloured as in 
          Fig. \ref{spectrum} according to their rotational
          band assignment. Experimental states are displayed 
          as black dots. Our identifications between theoretical
          and experimental states are shown by lines joining the
          states.} \label{experimental}	
\end{figure}

The theoretical energy spectrum of our model is plotted in Fig. 
\ref{spectrum}, where states from each rotational band are displayed in a 
different colour. The total energy of each state is 
\be
E = E_{\rm vib} + n_A\omega_A + n_F\omega_F + E_{\rm rot} \,,
\label{Etotal}
\ee
where $E_{\rm vib}$ is the underlying E-manifold energy, $n_A$ and
$n_F$ the number of A- and F-phonons, with $\omega_A = 12.05$ MeV
and $\omega_F = 6.55$ MeV their frequencies, and $E_{\rm rot}$ is the total
rotational energy given by, respectively, eqs. \eqref{Erot}, \eqref{ErotF} 
or \eqref{ErotF2} in the cases of no F-phonons, one F-phonon or two 
F-phonons. We plot $E$ against $J$. The spectrum is rather
dense and so we plot it again in Fig. \ref{experimental}. Here, each
spin/parity is considered separately and the figure also includes our
proposed identification of model states with experimental states. Finally we
tabulate the theoretical energies in Appendix A. The most important
results are discussed below.

States with small fixed spin ($J\leq4$) are generally ordered by their
vibrational energy $E_{\rm vib}$. Hence, the lowest-lying $0^+, 3^-$ and 
$4^+$ states all come from the ${\rm E}^0$ wavefunctions. They form the
tetrahedral ground-state rotational band. Extrapolating this band to higher 
spins, using our energy formula, gives $6^\pm, 7^-, 8^+$ and $9^\pm$
states. Of these, the natural parity states (having parity $(-1)^J$) 
have been observed at energies close to the predicted values, but their 
energies are not the lowest for those spins, because of band
crossing. This was noted earlier by Robson, who identified the 
first-excited $6^+$ state at 16.27 MeV as belonging to the 
tetrahedral ground-state band, although there is a lower $6^+$ state 
at 14.82 MeV.
  
The next band is the ${\rm E}^1$-band. Its lowest-energy states 
are a $2^+$ state and a somewhat higher $2^-$ state with predicted 
energies close to those of the lowest observed states with these spin/parities. 
Experimentally, the energy splitting is $1.95$ MeV, compared with 
$1.82$ MeV in our model. The splitting is caused by the difference 
in vibrational energy due to tunnelling in the E-manifold. At higher
spin, the ${\rm E}^1$-band has the first-excited $4^+$ state and the 
lowest predicted $4^-$ state, and the lowest $5^+$ and $5^-$ states, 
always with the same splitting between the parity doubles. The 
lowest-energy $6^+$ state also appears to be in this band, although its 
observed energy of 14.82 MeV is less than what our model predicts. The 
lowest observed $8^+$ state is in this band too while the F-band 
gives rise to the lowest-energy $7^-$ state.

From spin 6 upwards, no unnatural parity states have yet been observed,
but our model predicts several $6^-$, $7^+$ etc. states
above 17 MeV. Each band has a selection of these, and
their predicted energies are shown in the figures.  

The ${\rm E}^2$-band is interesting, because in our model, the
first-excited $0^+$ state at 6.05 MeV is interpreted as belonging to this band.
Because of the representation decomposition $E^2 = A_1 \oplus E$, this
band combines the spin/parities occurring in the ground-state ${\rm E}^0$-band 
with those in the ${\rm E}^1$-band, so it has states with spin/parities $0^+$,
$3^-$ and $4^+$, and also $2^\pm$ and $4^\pm$. The predicted $3^-$ state can be
identified with the fourth-excited $3^-$ state at $15.41$ MeV. This
unexpectedly high energy occurs because of the relatively large vibrational
energy of 10.67 MeV, and shows the
importance of including the effect of tunnelling. Models which treat this
state as a simple rotational excitation of the $0^+$ state at 6.05
MeV give it an energy of 12 MeV or less. However, such an approach 
leads to too many $3^-$ states with low energy. The $2^\pm$ states in
the ${\rm E}^2$-band match excited states with these
spin/parities, and the two $4^+$ states can be matched with the observed 
states either side of 15 MeV. The single $4^-$ state is predicted to 
lie above 18 MeV, close to where a couple of such states are observed.

The states in the F-band match experimental states quite well. For
these, there is a significant contribution from the Coriolis
effect. The band has the lowest $1^-$ state and the first-excited $2^+$ 
state, then close together the lowest $3^+$ state observed at 11.08
MeV and the first-excited $3^-$ state at 11.60 MeV. The $3^+$ state 
is predicted to have energy higher than the $3^-$ state just 
from the difference in vibrational energy, but the Coriolis correction
reverses this and makes the $3^+$ state lower, to match the data. The
$4^+$ state matches an observed state, and the $F$-band also has a 
nearby $4^-$ state. So from the ${\rm E}^1$-band and F-band two
$4^-$ states are predicted below 15 MeV, at 12.69 MeV and 13.84 MeV 
respectively, but just one is observed at
14.30 MeV. This is the first serious difference between the
predictions of our model and what is observed. There is also some
experimental uncertainty here \cite{O16Exp}, although a state is clearly seen
according to Kemper {\it et al.} \cite{Kem}, and probably has
unnatural parity. Our model clearly predicts two $4^-$ states in the 
12-15 MeV range, although they could overlap. The model of Bijker 
and Iachello makes a similar prediction \cite{TetVib}. In the F-band there 
are two $5^-$ states and one $5^+$ state. Further spin
5 states are predicted below $20$ MeV, arising from the 
${\rm E} \times {\rm F}$ band and ${\rm E}^2$-band. Compared to what
has been seen in experiments, 
we predict two additional $5^+$ states and three additional $5^-$
states. The observed $5^-$ peak at $14.66$ MeV is unusually broad,
and our model suggests it could arise from overlapping states in 
the ${\rm E}^1$- and F-bands with similar energies.

Let us now consider the higher-energy $0^+$ states. It seems agreed by
many authors that there is no $0^+$ state at 11.26 MeV. (It is recorded 
in the experimental tables, but observed with a weak signal in just one
experiment \cite{BM}.) There are clearly observed $0^+$ states at
12.05, 14.03 and 15.10 MeV. In our model we can match these to states 
in the ${\rm F}^2$-, ${\rm E}^3$- and A-bands. The predicted energies
for the first two of these are 13.11 and 14.89 MeV, so it is most likely
that the A-band state is at 12.05 MeV. This is the reason we have
calibrated the A-mode, breather frequency to be 12.05 MeV, close to 
the value of 11.4 MeV estimated from the breather state of 
Carbon-12, as mentioned in the Introduction. 

The ${\rm E}^3$-band also contains a $0^-$ state (because of the $A_2$ irrep
in the decomposition of $E^3$), whose energy is predicted to be 16.35 MeV. 
This is far higher than the observed energy of 10.96 MeV for the lowest $0^-$
state, and is the worst prediction of our model. This problem was noted 
previously \cite{HKM}. The wavefunction is shown in 
Fig. \ref{wavefunctions} (top right), and its energy is rather
sensitive to the form of the potential on the E-manifold, 
away from the tetrahedron and square configurations where this 
wavefunction has to vanish. Possibly the energy can be lowered by 
adjusting the potential, but that would change all other energies, and 
we have not investigated the matter in detail.

The ${\rm E} \times {\rm F}$ band decomposes into $F_1$ and $F_2$ 
subbands. The states in the $F_2$ subband have the same spin/parities
as those in the F-band but somewhat higher energies; the states
in the $F_1$ subband have reversed parities. We will not describe in
detail the states in this and in higher bands. There are a number of 
states with spins up to 5, below 20 MeV, that can be roughly matched to
observed states, as shown in Figs. \ref{spectrum} and 
\ref{experimental}. In particular, these bands give about the right
number of $2^+$ and $4^+$ states to match the data. A number 
of $3^-$ states are predicted between 16 and 20 MeV, but none have 
so far been observed. A few $4^-$ states in the same energy range
are also predicted.

$3^+$ states are an important check for our model as only three are observed 
lying below 20 MeV, at 11.08, 15.78 and 16.82 MeV. The third of these
has uncertain spin and isospin. In our model, these three states arise from the 
F-band (one) and ${\rm E} \times {\rm F}$ band (two), at energies
11.39, 15.43 and 16.52 MeV. This indicates that the observed state at
16.82 MeV is definitely a $3^+$ state with isospin 0. The 
${\rm E} \times {\rm F}$ band also contains a single 
$1^+$ state at 12.79 MeV, fairly close in energy to the observed state 
at 13.66 MeV, the only such state known. The next $1^+$ state in our 
model is in the ${\rm E}^2 \times {\rm F}$ band and lies at 
18.57 MeV. These rare spin states are especially
important for comparing models. There are fifteen experimentally 
observed $2^+$ states below 20 MeV. Hence it is difficult to verify 
(or falsify) a model using only $2^+$ states, provided the model has 
lots of them. However an abundance of low-lying $1^+$ or $3^+$ states 
would be a significant failing.

\begin{figure}
	\begin{center}
		\includegraphics[width=0.9\textwidth]{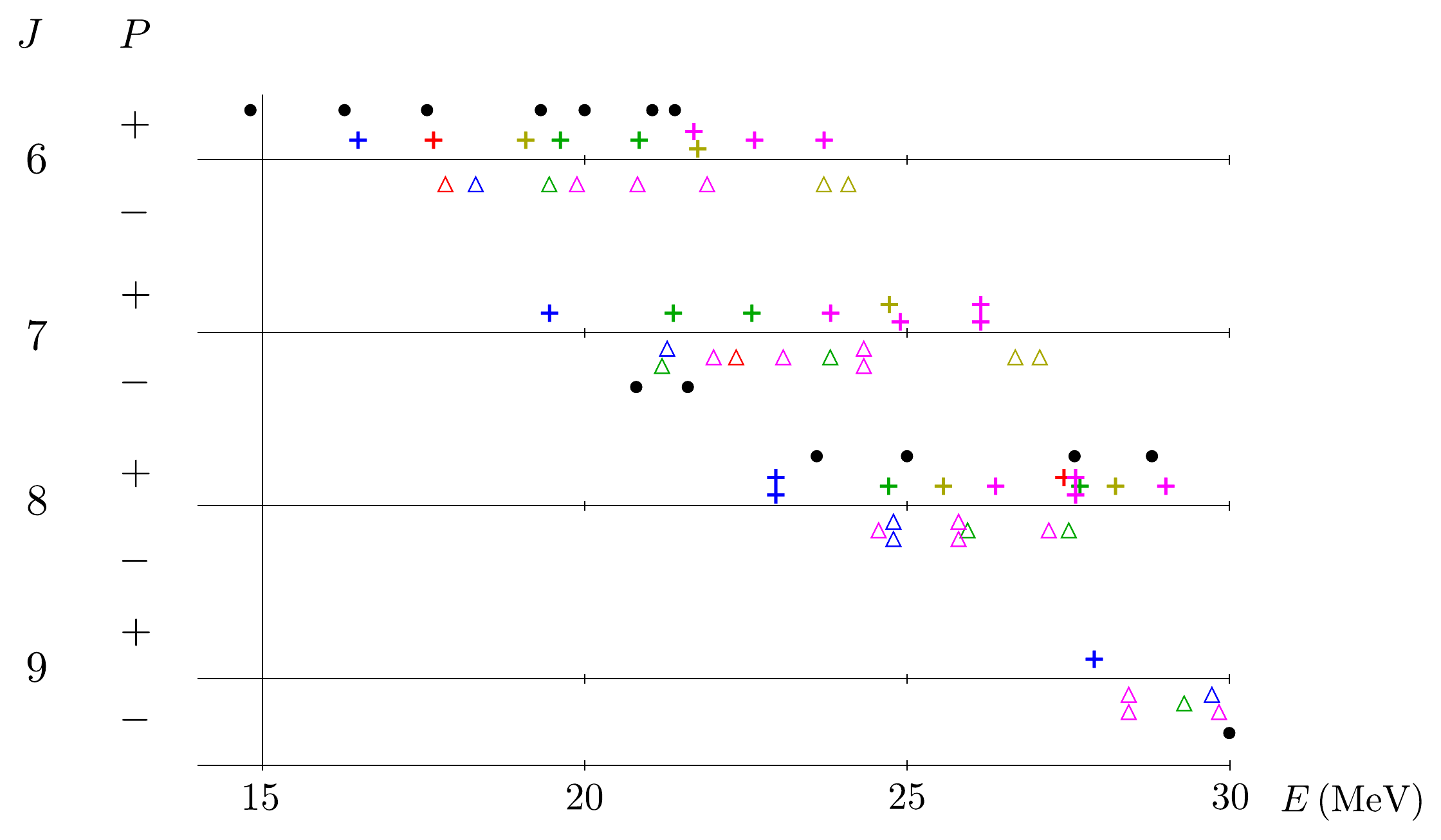} 
	\end{center}
	\caption{Comparison between theoretical and experimental
          data for high-spin states with energies above 15 MeV, using
          the notation of Fig. \ref{experimental}. Only states from 
          the ${\rm E}^0$-, ${\rm E}^1$-, ${\rm E}^2$-, F- and 
          ${\rm E} \times {\rm F}$ bands are displayed.} 
\label{highspin}	
\end{figure}

The most interesting states with energy greater than 20 MeV are those
with spin 6 or more, as some of these have been discovered recently 
\cite{Fre}, and do not appear in the tables \cite{O16Exp}. In fact, 
there has been significant debate within the literature about 
high-spin states of Oxygen-16. For example, Sanders \textit{et al.}
\cite{SMP} claimed to find an $8^+$ state between $22$ and $23$
MeV. Despite some effort in the years following, the state's existence
was never corroborated by others \cite{Ames, RAZ} and further, no
other $8^+$ state was discovered in this region. More recently, 
Freer \textit{et al.} found at least three $8^+$ states between 23 MeV
and 30 MeV by studying Beryllium-8 decay channels
\cite{Fre}. These authors also discovered a $6^+$ state at 21.2 MeV,
close in energy to other $6^+$ states at 21.4 MeV and 21.6 MeV seen in
the experiments of \cite{Fre2} and \cite{Ames} respectively. We will
assume that these three states are all from a single broad
resonance. There are two established $7^-$ states at 20.86 MeV and
21.62 MeV, and probably more at higher energy. There is also evidence 
of a $9^-$ state at around 30 MeV although its exact energy is 
uncertain \cite{RAZ}. Experimental work at these energies is difficult and
we expect more states will be discovered as experimental techniques
continue to improve. Fig. \ref{highspin} shows the states predicted 
by our model, compared with the known experimental data, for
$6^+$, $7^-$, $8^+$ and $9^-$ states. The states shown are from the 
${\rm E}^0$-, ${\rm E}^1$-, ${\rm E}^2$-, F- and 
${\rm E} \times {\rm F}$ bands, although further $6^+$ and
$7^-$ states with energy below 30 MeV are likely to arise from higher
bands. The model predicts a similar number of unnatural parity states,
but none have yet been observed. Our high-spin spectrum is dense with
states and includes numerous degeneracies. This is characteristic of
any cluster model. Since states at this energy often appear as broad
resonances, it may be difficult to experimentally
distinguish individual states which are nearby in energy. Regardless,
our model predicts a high-spin spectrum just as dense and complex as
what is seen at lower energies.

Overall, the model successfully describes approximately 70 states in 
the observed spectrum of Oxygen-16, with most of the energy
predictions matching measured energies within about 1 MeV. 
Exceptionally, the predicted energy for the lowest $0^-$ state is
about 5 MeV too high. A number of $3^-$ and $4^-$ states just below 20 MeV are
predicted, and numerous unnatural parity states with high spin should
exist, starting with a $6^-$ state in the ground-state band between 17
MeV and 18 MeV. Perhaps our most important prediction is the existence 
of a further $4^-$ state around 13 MeV. Finding such a state would 
help confirm the tetrahedral cluster model approach to the Oxygen-16 
nucleus, and the importance of considering the global structure of 
the E-manifold and the effect of tunnelling.

\section*{Acknowledgements}

This work has been partially supported by STFC consolidated grant
ST/P000681/1. CJH is supported by The Leverhulme Trust as an Early 
Careers Fellow. CK was supported by an STFC studentship.

\section*{Appendix A}

Table 1 lists all the quantum states below 20 MeV predicted by
our model, and also those with spin 6 or more having energy less than 
30 MeV. They are grouped into tetrahedral rovibrational bands.

\begin{tabularx}{\linewidth}{ l | l l l l l } 
	
		\hline \hline
		Band & $J^P$ & Vib. wvfn. & Vib. ener. &
                $E_\text{rot}$ & $E$ \\ \hline
		
                ${\rm E}^0$ & $0^+$ & $\psi^+_{T0}$ & 0 & 0 & 0 \\
		& $3^-$ & $\psi^-_{S0}$ & 0.18 & 6.12 & 6.30 \\
		& $4^+$ & $\psi^+_{T0}$ & 0 & 9.62 & 9.62 \\
		& $6^+$ & $\psi^+_{T0}$ & 0 & 17.66 & 17.66 \\
		& $6^-$ & $\psi^-_{S0}$ & 0.18 & 17.66 & 17.84 \\
		& $7^-$ & $\psi^-_{S0}$ & 0.18 & 22.17 & 22.35 \\
		& $8^+$ & $\psi^+_{T0}$ & 0 & 27.44 & 27.44  \\ \hline
		
		${\rm E}^1$ & $2^+$ & $u^+_1 \pm v_1^+$ & 3.45 & 2.54 & 5.99 \\
		& $2^-$ & $u^-_1 \pm v_1^-$ & 5.27 & 2.54 & 7.81 \\
		& $4^+$ & $u^+_1 \pm v_1^+$ & 3.45 & 7.42 & 10.87 \\
		& $4^-$ & $u^-_1 \pm v_1^-$ & 5.27 & 7.42 & 12.69 \\
		& $5^+$ & $u^+_1 \pm v_1^+$ & 3.45 & 10.21 & 13.66 \\
		& $5^-$ & $u^-_1 \pm v_1^-$ & 5.27 & 10.21 & 15.48 \\
		& $6^+$ & $u^+_1 \pm v_1^+$ & 3.45 & 13.04 & 16.49 \\
		& $6^-$ & $u^-_1 \pm v_1^-$ & 5.27 & 13.04 & 18.31 \\
		& $7^+$ & $u^+_1 \pm v_1^+$ & 3.45 & 16.01 & 19.46 \\
		& $7^-$ & $u^-_1 \pm v_1^-$ & 5.27 & 16.01 & 21.28 \\
		& $8^+$ & $u^+_1 \pm v_1^+$ & 3.45 & 19.52 & 22.97 \\
		& $8^+$ & $u^+_1 \pm v_1^+$ & 3.45 & 19.52 & 22.97 \\
		& $8^-$ & $u^-_1 \pm v_1^-$ & 5.27 & 19.52 & 24.79 \\
		& $8^-$ & $u^-_1 \pm v_1^-$ & 5.27 & 19.52 & 24.79 \\
		& $9^+$ & $u^+_1 \pm v_1^+$ & 3.45 & 24.46 & 27.91 \\
		& $9^-$ & $u^-_1 \pm v_1^-$ & 5.27 & 24.46 & 29.73 \\ \hline
		
		${\rm E}^2$ & $0^+$ & $\psi^+_{T2}$ & 6.05 & 0 & 6.05 \\
		& $2^+$ & $u^+_2 \pm v_2^+$ & 8.72 & 2.54 & 11.26 \\
		& $2^-$ & $u^-_2 \pm v_2^-$ & 11.05 & 2.54 & 13.59 \\
		& $3^-$ & $\psi^-_{S2}$ & 10.67 & 4.80 & 15.47 \\
		& $4^+$ & $\psi^+_{T2}$ & 6.05 & 7.42 & 13.47 \\
		& $4^+$ & $u^+_2 \pm v_2^+$ & 8.72 & 7.42 & 16.14 \\
		& $4^-$ & $u^-_2 \pm v_2^-$ & 11.05 & 7.42 & 18.47 \\
		& $5^+$ & $u^+_2 \pm v_2^+$ & 8.72 & 10.21 & 18.93 \\
		& $5^-$ & $u^-_2 \pm v_2^-$ & 11.05 & 10.21 & 21.26 \\
		& $6^+$ & $\psi^+_{T2}$ & 6.05 & 13.04 & 19.09 \\
		& $6^-$ & $\psi^-_{S2}$ & 10.67 & 13.04 & 23.71 \\
		& $6^+$ & $u^+_2 \pm v_2^+$ & 8.72 & 13.04 & 21.76 \\
		& $6^-$ & $u^-_2 \pm v_2^-$ & 11.05 & 13.04 & 24.09 \\
		& $7^+$ & $u^+_2 \pm v_2^+$ & 8.72 & 16.01 & 24.73 \\
		& $7^-$ & $u^-_2 \pm v_2^-$ & 11.05 & 16.01 & 27.06 \\
		& $7^-$ & $\psi^-_{S2}$ & 10.67 & 16.01 & 26.68 \\
		& $8^+$ & $u^+_2 \pm v_2^+$ & 8.72 & 19.52 & 28.24 \\
		& $8^+$ & $u^+_2 \pm v_2^+$ & 8.72 & 19.52 & 28.24 \\
		& $8^+$ & $\psi^+_{T2}$ & 6.05 & 19.52 & 25.57 \\ \hline
		
		${\rm E}^3$ & $0^+$ & $\psi^+_{T3}$ & 14.89 & 0 & 14.89 \\
		& $0^-$ & $\psi^-_{T3}$ & 16.35 & 0 & 16.35 \\
		& $2^+$ & $u^+_3 \pm v_3^+$ & 12.22 & 2.54 & 14.76 \\
		& $2^-$ & $u^-_3 \pm v_3^-$ & 16.92 & 2.54 & 19.46 \\
		& $3^+$ & $\psi^+_{S3}$ & 12.57 & 4.80 & 17.37 \\
		& $4^+$ & $u^+_3 \pm v_3^+$ & 12.22 & 7.42 & 19.64 \\ \hline
		
		${\rm E}^4$ & $0^+$ & $\psi^+_{T4}$ & 18.78 & 0 
                & 18.78 \\ \hline
		
		${\rm F}$ & $1^-$ & $\psi^+_{T0}$ & 6.55 & 0.57 & 7.12 \\
		& $2^+$ & $\psi^-_{S0}$ & 6.73 & 2.93 & 9.66 \\
		& $3^+$ & $\psi^-_{S0}$ & 6.73 & 4.66 & 11.39 \\
		& $3^-$ & $\psi^+_{T0}$ & 6.55 & 5.36 & 11.91 \\
		& $4^+$ & $\psi^-_{S0}$ & 6.73 & 6.58 & 13.32 \\
		& $4^-$ & $\psi^+_{T0}$ & 6.55 & 7.28 & 13.84 \\
		& $5^+$ & $\psi^-_{S0}$ & 6.73 & 11.12 & 17.85 \\
		& $5^-$ & $\psi^+_{T0}$ & 6.55 & 9.19 & 15.75 \\
		& $5^-$ & $\psi^+_{T0}$ & 6.55 & 11.12 & 17.67 \\
		& $6^+$ & $\psi^-_{S0}$ & 6.73 & 14.12 & 20.85 \\
		& $6^+$ & $\psi^-_{S0}$ & 6.73 & 12.90 & 19.63 \\
		& $6^-$ & $\psi^+_{T0}$ & 6.55 & 12.90 & 19.45 \\
		& $7^+$ & $\psi^-_{S0}$ & 6.73 & 14.64 & 21.38 \\
		& $7^+$ & $\psi^-_{S0}$ & 6.73 & 15.86 & 22.60 \\
		& $7^-$ & $\psi^+_{T0}$ & 6.55 & 14.64 & 21.20 \\
		& $7^-$ & $\psi^+_{T0}$ & 6.55 & 17.27 & 23.81 \\
		& $8^+$ & $\psi^-_{S0}$ & 6.73 & 17.98 & 24.72 \\
		& $8^+$ & $\psi^-_{S0}$ & 6.73 & 20.96 & 27.69 \\
		& $8^-$ & $\psi^+_{T0}$ & 6.55 & 19.38 & 25.94 \\
		& $8^-$ & $\psi^+_{T0}$ & 6.55 & 20.96 & 27.51 \\
		& $9^-$ & $\psi^+_{T0}$ & 6.55 & 22.74 & 29.30 \\ \hline
		
		${\rm A}$ & $0^+$ & $\psi^+_{T0}$ & 12.05 & 0 & 12.05 \\ 
		& $3^-$ & $\psi^-_{S0}$ & 12.23 & 4.80 & 17.03 \\ 
		& $4^+$ & $\psi^+_{T0}$ & 12.05 & 7.42 & 19.47 \\  \hline
		
                ${\rm E} \times {\rm F}$ & $1^+$ & $u^-_1 \pm v_1^-$ 
                & 11.82 & 0.97 & 12.79 \\
		& $1^-$ & $u^+_1 \pm v_1^+$ & 10.00 & 0.97 & 10.97 \\
		& $2^+$ & $u^-_1 \pm v_1^-$ & 11.82 & 2.12 & 13.94 \\
		& $2^-$ & $u^+_1 \pm v_1^+$ & 10.00 & 2.12 & 12.12 \\
		& $3^+$ & $u^-_1 \pm v_1^-$ & 11.82 & 3.61 & 15.43 \\
		& $3^+$ & $u^-_1 \pm v_1^-$ & 11.82 & 4.70 & 16.52 \\
		& $3^-$ & $u^+_1 \pm v_1^+$ & 10.00 & 3.61 & 13.61 \\
		& $3^-$ & $u^+_1 \pm v_1^+$ & 10.00 & 4.70 & 14.70 \\
		& $4^+$ & $u^-_1 \pm v_1^-$ & 11.82 & 6.30 & 18.12 \\
		& $4^+$ & $u^-_1 \pm v_1^-$ & 11.82 & 7.08 & 18.90 \\
		& $4^-$ & $u^+_1 \pm v_1^+$ & 10.00 & 6.30 & 16.30 \\
		& $4^-$ & $u^+_1 \pm v_1^+$ & 10.00 & 7.08 & 17.08 \\
		& $5^+$ & $u^-_1 \pm v_1^-$ & 11.82 & 7.80 & 19.63 \\
		& $5^+$ & $u^-_1 \pm v_1^-$ & 11.82 & 8.58 & 20.41 \\
		& $5^+$ & $u^-_1 \pm v_1^-$ & 11.82 & 9.52 & 21.34 \\
		& $5^-$ & $u^+_1 \pm v_1^+$ & 10.00 & 7.80 & 17.81 \\
		& $5^-$ & $u^+_1 \pm v_1^+$ & 10.00 & 8.58 & 18.59 \\
		& $5^-$ & $u^+_1 \pm v_1^+$ & 10.00 & 9.52 & 19.52 \\
		& $6^+$ & $u^-_1 \pm v_1^-$ & 11.82 & 9.87 & 21.70 \\
		& $6^+$ & $u^-_1 \pm v_1^-$ & 11.82 & 10.81 & 22.64 \\
		& $6^+$ & $u^-_1 \pm v_1^-$ & 11.82 & 11.90 & 23.72 \\
		& $6^-$ & $u^+_1 \pm v_1^+$ & 10.00 & 9.87 & 19.88 \\
		& $6^-$ & $u^+_1 \pm v_1^+$ & 10.00 & 10.81 & 20.82 \\
		& $6^-$ & $u^+_1 \pm v_1^+$ & 10.00 & 11.90 & 21.90 \\
		& $7^+$ & $u^-_1 \pm v_1^-$ & 11.82 & 11.99 & 23.82 \\
		& $7^+$ & $u^-_1 \pm v_1^-$ & 11.82 & 13.08 & 24.90 \\
		& $7^+$ & $u^-_1 \pm v_1^-$ & 11.82 & 14.33 & 26.15 \\
		& $7^+$ & $u^-_1 \pm v_1^-$ & 11.82 & 14.33 & 26.15 \\
		& $7^-$ & $u^+_1 \pm v_1^+$ & 10.00 & 11.99 & 22.00 \\
		& $7^-$ & $u^+_1 \pm v_1^+$ & 10.00 & 13.08 & 23.08 \\
		& $7^-$ & $u^+_1 \pm v_1^+$ & 10.00 & 14.33 & 24.33 \\
		& $7^-$ & $u^+_1 \pm v_1^+$ & 10.00 & 14.33 & 24.33 \\
		& $8^+$ & $u^-_1 \pm v_1^-$ & 11.82 & 14.55 & 26.38 \\
		& $8^+$ & $u^-_1 \pm v_1^-$ & 11.82 & 15.80 & 27.62 \\
		& $8^+$ & $u^-_1 \pm v_1^-$ & 11.82 & 15.80 & 27.62 \\
		& $8^+$ & $u^-_1 \pm v_1^-$ & 11.82 & 17.20 & 29.02 \\
		& $8^-$ & $u^+_1 \pm v_1^+$ & 10.00 & 14.55 & 24.56 \\
		& $8^-$ & $u^+_1 \pm v_1^+$ & 10.00 & 15.80 & 25.80 \\
		& $8^-$ & $u^+_1 \pm v_1^+$ & 10.00 & 15.80 & 25.80 \\
		& $8^-$ & $u^+_1 \pm v_1^+$ & 10.00 & 17.20 & 27.20 \\
		& $9^-$ & $u^+_1 \pm v_1^+$ & 10.00 & 18.43 & 28.44 \\
		& $9^-$ & $u^+_1 \pm v_1^+$ & 10.00 & 18.43 & 28.44 \\
		& $9^-$ & $u^+_1 \pm v_1^+$ & 10.00 & 19.84 & 29.84 \\ \hline
		
		${\rm E}^2 \times {\rm F}$ & $1^+$ & $u^-_2 \pm v_2^-$ 
                & 17.60 & 0.97 & 18.57 \\
		& $1^-$ & $u^+_2 \pm v_2^+$ & 15.27 & 0.97 & 16.24 \\
		& $1^-$ & $\psi^+_{T2}$ & 12.60 & 0.50 & 13.11 \\
		& $2^+$ & $\psi^-_{S2}$ & 17.22 & 2.59 & 19.81 \\
		& $2^+$ & $u^-_2 \pm v_2^-$ & 17.60 & 2.12 & 19.72 \\
		& $2^-$ & $u^+_2 \pm v_2^+$ & 15.27 & 2.12 & 17.39 \\
		& $3^-$ & $\psi^+_{T2}$ & 12.60 & 4.70 & 17.30 \\
		& $3^-$ & $u^+_2 \pm v_2^+$ & 15.27 & 3.61 & 18.88 \\
		& $3^-$ & $u^+_2 \pm v_2^+$ & 15.27 & 4.70 & 19.97 \\ \hline
				
                ${\rm F}^2$ & $0^+$ & $\psi^+_{T0}$ & 13.11 & 0 & 13.11 \\
		& $1^-$ & $\psi^-_{S0}$ & 13.29 & 0.56 & 13.85 \\
		& $2^+$ & $\psi^+_{T0}$ & 13.11 & 3.27 & 16.37 \\
		& $2^+$ & $\psi^+_{T0}$ & 13.11 & 1.71 & 14.82 \\
		& $2^-$ & $\psi^-_{S0}$ & 13.29 & 2.34 & 15.62 \\
		& $3^+$ & $\psi^+_{T0}$ & 13.11 & 4.14 & 17.24 \\
		& $3^-$ & $\psi^-_{S0}$ & 13.29 & 4.76 & 18.05 \\
		& $3^-$ & $\psi^-_{S0}$ & 13.29 & 4.20 & 17.49 \\
		& $4^+$ & $\psi^+_{T0}$ & 13.11 & 6.42 & 19.53 \\
		& $4^+$ & $\psi^+_{T0}$ & 13.11 & 5.27 & 18.38 \\
		& $4^-$ & $\psi^-_{S0}$ & 13.29 & 5.27 & 18.56 \\ \hline
		
                ${\rm E} \times {\rm A}$ & $2^+$ & $u^+_1 \pm v_1^+$ 
                & 15.50 & 2.54 & 18.04 \\ 
		& $2^-$ & $u^-_1 \pm v_1^-$ & 17.32 & 2.54 & 19.86 \\  \hline
		 
		${\rm E}^2 \times {\rm A}$ & $0^+$ & $\psi^+_{T2}$ 
                & 18.10 & 0 & 18.10 \\
		
		\hline
		
		\caption{Bands of predicted quantum states. We
                  tabulate for each state the spin/parity $J^P$, 
                  the underlying vibrational wavefunction on the 
                  E-manifold (displayed in Fig. \ref{wavefunctions}),
                  the vibrational energy $E_{\rm vib} + n_A\omega_A 
                  + n_F\omega_F$, the rotational energy $E_{\rm rot}$ 
                  and total energy $E$ (all in MeV).}

\end{tabularx}

\end{document}